\documentclass[%
superscriptaddress,
groupedaddress,
showpacs,preprintnumbers,
 amsmath,amssymb,
 aps,
pre,
floatfix,
12pt
 ]{revtex4-1}
\usepackage{physics}
\usepackage{xcolor}
\usepackage{graphicx}% Include figure files
\usepackage{dcolumn}% Align table columns on decimal point
\usepackage[mathscr]{euscript}
\usepackage{bm}% bold math
\usepackage{caption}
\usepackage{subcaption}
\usepackage[shortlabels]{enumitem}
\usepackage{amsmath}
\usepackage{amssymb}
\begin{document}

\title{Counterfactual Thermodynamics: Extracting work from a lack of macroscopic change}
\author{Sushrut Ghonge}
\affiliation{
 Department of Physics, University of Notre Dame, Notre Dame, Indiana 46556, USA}

\author{Dervis Can Vural}
\email{dvural@nd.edu}
\affiliation{
 Department of Physics, University of Notre Dame, Notre Dame, Indiana 46556, USA}

\begin{abstract}
A sudden change in the macroscopic parameters of a system will cause it to depart from equilibrium. In this paper we study how a lack of change can also inform of such a departure, and allow for work extraction. Potential events that are unrealized can provide information pertaining the microstates of the system, essentially playing the role of a passive Maxwell demon, thereby allowing one to infer details about the microstate probability distribution. Here, we first qualitatively argue that this effect is present and consequential in almost every physical system, but is ignored in the standard formulation of equilibrium statistical mechanics. Then, as a case study, we quantitatively investigate the local and global thermodynamic properties of an ideal gas placed in a fragile container that nevertheless, does not burst. This non-event indicates a departure from equilibrium and allows work extraction from the system. It also leads to corrections to the heat capacity of the gas.
\end{abstract}
\maketitle

\section{Introduction}
Intuition suggests that if the macroscopic parameters of a closed system are kept constant, the system will approachw thermodynamic equilibrium \cite{reiss2012methods}. This need not be the case. Here we study the thermodynamic properties of a simple system in which the \emph{absence of} macroscopic change constitutes a continuous stream of information which indicates a departure from equilibrium, and allows work extraction from the system.
%causing the system to go out of equilibrium.

Consider for example, a gas sampled from a thermal bath, in equilibrium, with temperature $T$ and placed in an isolated container such that neither the gas nor the container undergoes an observable macroscopic change during some time $\tau$. At this stage, the fact that the container has not burst or deformed or leaked so far, informs that no molecule above a critical energy has yet hit the walls of the container. Since the original Boltzmann distribution actually included such high energies, the statistical description of the gas must be modified as a function of $\tau$. In this example, a non-event informs us of a position and time dependent energy distribution. The lack of an equilibrium-disrupting event is indicating %leading to
a departure from equilibrium.

In similar vein, consider a molecule placed in a gas with which it can react explosively upon a sufficiently energetic collision. The lack of such an explosion indicates that there were no high-energy particles on the trajectory of the molecule. In other words, we can infer a region of low temperature - a cold trail - on the path of the explosive molecule.

Of course, similar qualitative arguments can be made more generally for other macroscopic non-events: Lack of chemical and nuclear reactions, evaporation and condensation, adsorption and desorption, dissolution and precipitation all occur when system constituents are within a specific energy window, and typically with higher likelihood near certain locations. Any system that has a \emph{potential} to undergo such processes starts revealing its microstates every moment this potential is not realized. The longer nothing happens, the more information one gathers about the system, and the further away the system has departed from its original state of equilibrium.

This odd conclusion does not stem from the semantics of what is meant by thermodynamic equilibrium \cite{pitowsky2006definition,werndl2015rethinking}. It has concrete physical consequences. Since local information can be used to extract work \cite{puglisi2017clausius,puglisi2017temperature,rupprecht2019maxwell,koski2015chip,aaberg2013truly, parrondo2015thermodynamics}, we will show that a lack of change in a system can also allow work extraction.

We will start with a specific system -- an ideal gas placed in a non-bursting fragile container --, and quantify the local deviations from the Gibbs-Boltzmann distribution, as well modifications to global thermodynamic properties of a simple system, as a function of the time $\tau$ during which the container remains intact.  We will show that a local temperature variation, as inferred by the non-event, as well as the evolving difference in global temperatures between the container and the original thermal bath, can be used to generate some amount of work, $W(\tau)$, which we will quantify.

Interestingly, based on the duration of the non-event, we can also make retrodictions and predictions (similar to \cite{rupprecht2018limits,rupprecht2019enhancing}) about the past and the future of the local temperature distribution in the container. Furthermore, we also show how the heat capacity $C_v(T)=dE/dT$ of the gas must also be modified in light of the non-event.

Of course, none of this violates any fundamental laws of physics. %A system can be driven out of equilibrium
The statistical distribution of a system in phase space can be changed 
by performing measurements whose outcome restricts the space of microstates available to the system \cite{rupprecht2019maxwell, koski2015chip,hartmann2004local,brites2012thermometry, muller2019information, naghiloo2018information}, thereby turning the equilibrium state into a fluctuation state \cite{hickman2016temperature,mishin2015thermodynamic,bertini2015macroscopic,qian2001relative}. Such deviations from equilibrium were first quantified by the fluctuation theorems \cite{jarzynski1997equilibrium,jarzynski1997nonequilibrium,crooks1999entropy,kurchan1998fluctuation,seifert2005entropy,evans1993probability,wang2002experimental,evans1994equilibrium,evans2002fluctuation,gallavotti1995dynamical}, and have interesting consequences such as second law violation. They have since been generalized to quantum systems and systems with feedback control \cite{talkner2007tasaki,kurchan2000quantum,sagawa2008second,sagawa2010generalized,sagawa2012nonequilibrium,callen1951irreversibility}. The study of the fluctuations of a system and the information contained in them forms the subject of stochastic thermodynamics \cite{seifert2012stochastic,seifert2008stochastic,ciliberto2017experiments,bechhoefer2020stochastic,seifert2019stochastic,jarzynski2011equalities,klages2013nonequilibrium}. Like almost all the results in this field, our predictions, while valid for systems of any size, can be most easily observed in finite particle systems.

As we conclude our introduction, we should point out that whether a measurement of a microstate \emph{causes} a deviation from equilibrium or simply \emph{informs} of such a deviation has been an issue of considerable debate \cite{shalizi2004backwards,jaynes1990probability,kyburg1987basic,sklar1995physics}. Depending on one's philosophical inclinations the present paper can be read in two ways. A Bayesian who thinks that an update in our subjective knowledge of a system causes an update in its entropy might interpret the local energy fluctuations as ``deviations from equilibrium \emph{caused} by a lack of change''. In contrast, a frequentist will treat the container of gas as a member of an imaginary ensemble of similar containers, and will view the non-event as a revelation of what kind of a container was ``picked'' from this ensemble. In this interpretation the local energy fluctuations were already present in the container and the lack of change \emph{informs} the observer of their presence. While we find merit in both interpretations, in the present context we prefer to leave it aside, and anchor our discussion solely on concrete physical observables such as energy distribution, heat capacity and work extraction.

\section{Model}

%---------FIGURE 1-------------%
\begin{figure}
    \includegraphics[width=0.7\linewidth]{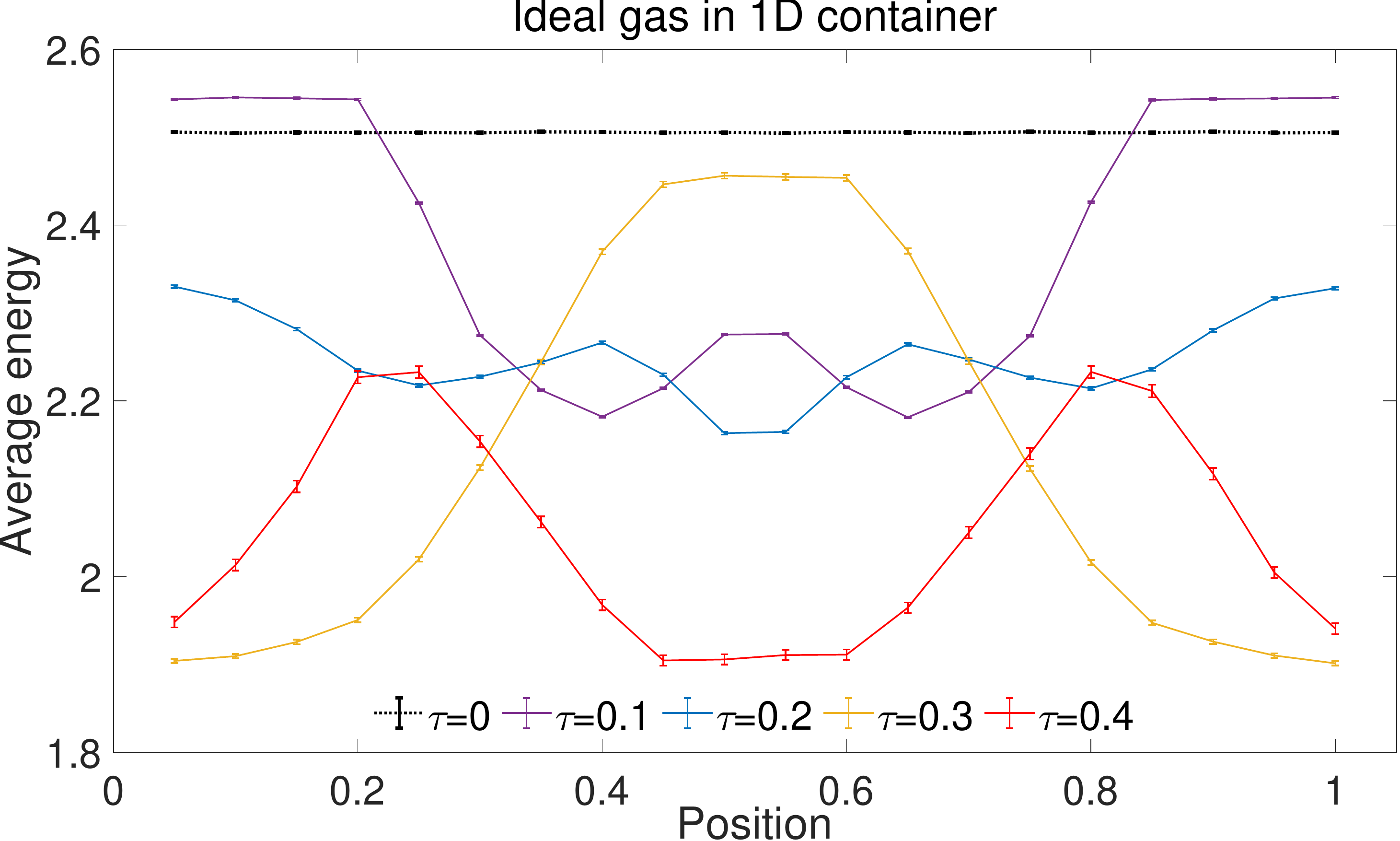}
    \caption{Inferred local mean energy at $t=\tau$, for a 1D fragile container. Also see corresponding videos included as supplemental material. $N=100, v_{c}=2$, $k_BT=L=m=1$ and the container domain is divided into 20 bins. The error bars show the standard error for $10^7$ Monte Carlo samples.}
    \label{fig:local1d}
\end{figure}
%---------FIGURE 1-------------%

Suppose that $N$ ideal-gas particles are sampled from a heat bath of temperature $T$ and placed in an isolated cubical container of size $L$. Suppose further that the container walls undergo a macroscopically visible change when hit by a particle whose normal velocity component is greater than a critical velocity, $v_c$. We refer to this event as a detection. Protocols for detecting high energy particles using walls of a container have been discussed in the context of explosions \cite{nowakowski2002thermal}. Chemical reactions caused by high velocity particles, including ignition and explosions, have also been studied extensively \cite{kramers1940brownian,van1987intrinsic,janssen1989elimination,hanggi1990reaction}.

Let $\tau$ be a time period of no detection and $E(\vec{r},t)$ be the energy density in an infinitesimal volume around $\vec{r}$ at time $t$. We use the Bayes' theorem to infer the probability that the energy density at $\vec{r}$ at time $t$ is $E(\vec{r},t)$,
\begin{equation}\label{bayesenergy}
    P(E;\vec{r},t | \tau)=P(\tau | E;\vec{r},t) P_{a}(E;\vec{r})/P(\tau).
\end{equation}
Here, $P(\tau)=\int \mathrm{d}E P(\tau | E;\vec{r},t) P_{a}(E;\vec{r})$ is an integral over all possible $E(\vec{r},t)$, and the \textit{a-priori} probability distribution, $P_a(E;\vec{r})$ is defined below.

For simplicity, we now consider particles in a one-dimensional box. The following equations can be easily generalized to higher dimensions. The \textit{a-priori} probability that a single particle in 1D has energy $E$ and is at a point $x$ is
\begin{equation}\label{singleapriori}
    P_{a,1}(E,x)\mathrm{d}E\mathrm{d}x= \exp(-E/kT)(\mathrm{d}x/L)(\mathrm{d}E/\sqrt{\pi kTE}).
\end{equation}

Whereas for $N$ particles with energies and positions $\{E_i,x_i\}$, the \textit{a-priori} probability density of total energy density $E$ at position $x$ is, 
\begin{equation}\label{apriori}
    P_{a,N}(E,x)=\int \textstyle{\prod_{i=1}^N} (\mathrm{d}E_i \mathrm{d}x_i) P_{a,1}(E_i,x_i) \delta\big[E-\textstyle{\sum_{j=1}^N E_j}\delta(x-x_j)\big].
\end{equation}
The summation over $j$ inside the $\delta$ function adds the energies of all particles that are close to $x$. The integral of the product over the index $i$ sums over all possible combinations of the positions ($x_i$) and energies ($E_i$) weighted by the corresponding probability density.

In a 1D container, a detection can occur at either at $0$ or $L$. For this to happen before some time $\tau$, there must be a particle with a velocity greater than $v_c$ and that particle should reach one of the walls before time $\tau$. Therefore, the conditional probability of no detection before $\tau$ given the energies and positions of the $N$ particles is,
\begin{align}\label{conditionalptau}
P_{N}(\tau |\{ E_i, x_i\},t) &= \textstyle{\prod_{i=1}^N} \theta (E_i-E_c) \eta(E_i,x_i)\\
\eta(E_i;x_i) &= \frac{1}{2}\big[\theta(t\!-\!\tau\!-\!x_i\textstyle\sqrt{\frac{2m}{E_i}}) + \theta(t\!-\!\tau\!-\!(L\!-\!x_i)\textstyle \sqrt{\frac{2m}{E_i}})\big]\nonumber
\end{align}
where $\theta$ is the unit step function, and $E_c=mv_c^2/2$. The two step functions in time are for two possible velocities, $\pm v_i$ that would lead to detection at $x=L$ and $x=0$. The total energy density as a function of position and time, $E(x,t)$, can easily be obtained from particle energies and positions, $\{E_i,x_i\}$, by binning --
    $E(x,t) = \sum_{i=1}^N E_i(t) \delta(x-x_i(t))$.

We can then obtain $P_N(\tau)$ by integrating over all possible energies and positions of the particles,
\begin{equation}\label{ptau}
    P_{N}(\tau)=\int \textstyle{\prod_{i=1}^N} (\mathrm{d}E_i \mathrm{d}x_i) P_{a,1}(E_i,x_i) P_{N}(\tau |\{ E_i, x_i\},t).
    %P_{N}(\tau)=\int\mathcal{D}\{E(x,t)\} \int\mathrm{d}x P_{N}(\tau | E(x,t))P_{a,N}(E,x).
\end{equation}
2D and 3D analogs of the above equations can be obtained by the same procedure. 
We can now substitute the expressions for $P_N(\tau)$, $P_{a,N}(E,x)$ and $P_{N}(\tau |E;x,t)$ (obtained by binning) into eq. \ref{bayesenergy} to obtain the inferred local energy distribution, which we discuss in the following section.

%---------FIGURE 2,3-------------%
\begin{figure}
    \includegraphics[width=\linewidth]{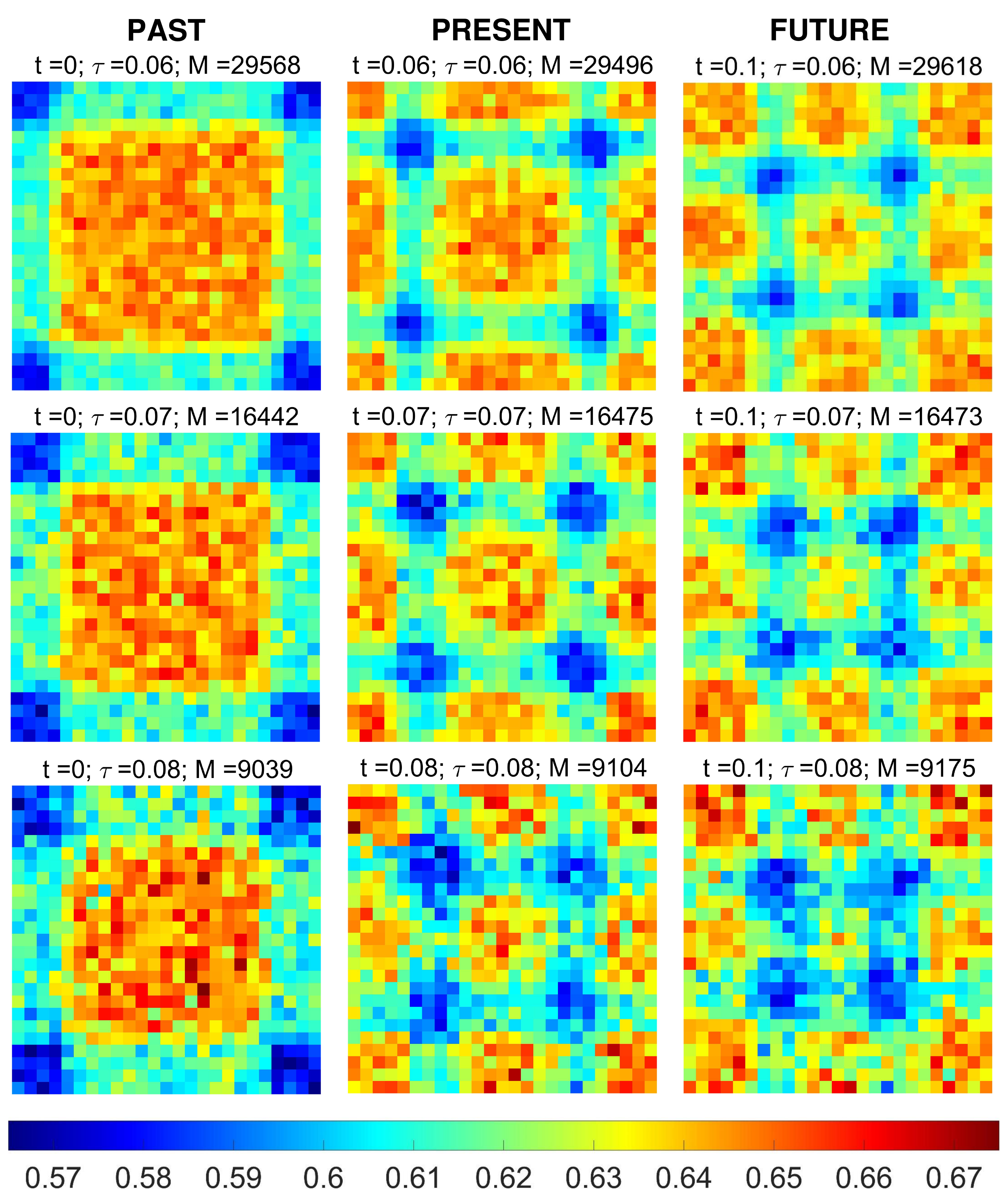}
    \caption{Heat maps of the local mean energy of an ideal gas in a 2D fragile container. Also see corresponding videos included as supplemental material. $\tau$ is varied along rows. Different columns show the inferred state of the system at different times, $t$ at present time $\tau$. The total number of Monte-Carlo trials is $10^6$, $k_BT=1$, $v_c=2, N=500$ and the box is divided into 625 bins in all cases. $M$ is the number of trials in which there was no detection until $\tau$. $k_B=L=m=1$. }
    \label{fig:localfig2d}
\end{figure}
\section{Variations in local mean energy} 
The probability distributions in eqns. \ref{apriori}, \ref{conditionalptau} and \ref{ptau} are then substituted into eqn.\ref{bayesenergy} to yield an energy distribution $P(E(\vec{r},t) | \tau)$ at every location $\vec{r}$. In Fig.\ref{fig:local1d} and Fig.\ref{fig:localfig2d} we use this  distribution to plot the local mean energy $\langle E(\vec{r})\rangle$ for various values of $\tau$. We can interpret this as the temperature variations within the container, as local mean energy is a convenient proxy for local temperature \cite{ghonge2018temperature,puglisi2017temperature}.

We evaluate the integrals in eqns.\ref{apriori}-\ref{ptau}, and their respective 2D analogs by Monte-Carlo integration. To do so, we first generate $10^6$ sets of initial positions and velocities with the \emph{a priori} distributions for $N$ particles. We then select the configurations that lead to detection times greater than $\tau$ and calculate the local mean energy of those configurations.

\begin{figure}
    \begin{subfigure}[b]{0.48\textwidth}
    \includegraphics[width=\linewidth]{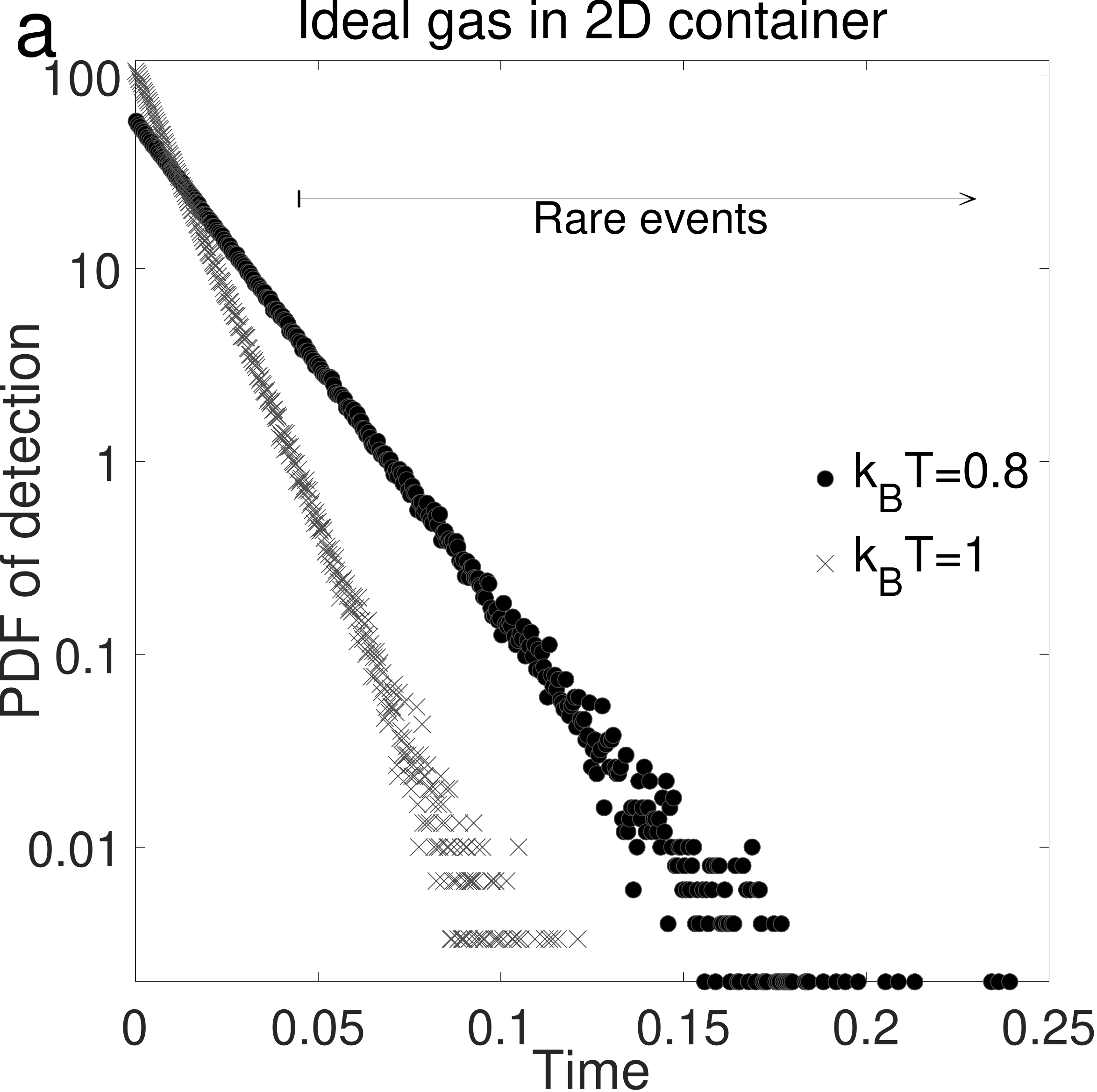}
    \end{subfigure}
    \hfill
    \begin{subfigure}[b]{0.48\textwidth}
    \includegraphics[width=\linewidth]{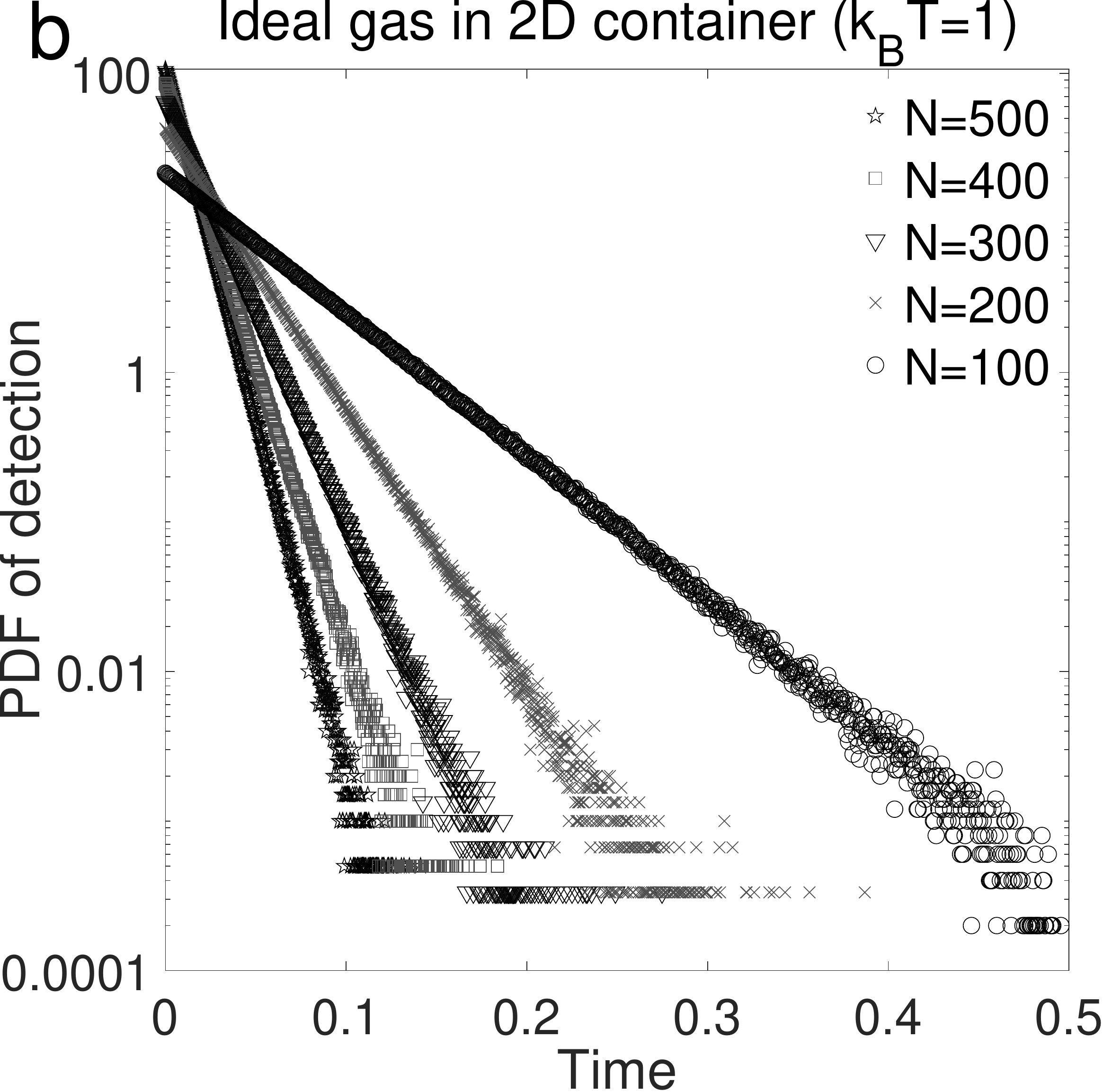}
    \end{subfigure}
    \par \medskip
    \begin{subfigure}[b]{0.32\textwidth}
    \includegraphics[width=\linewidth]{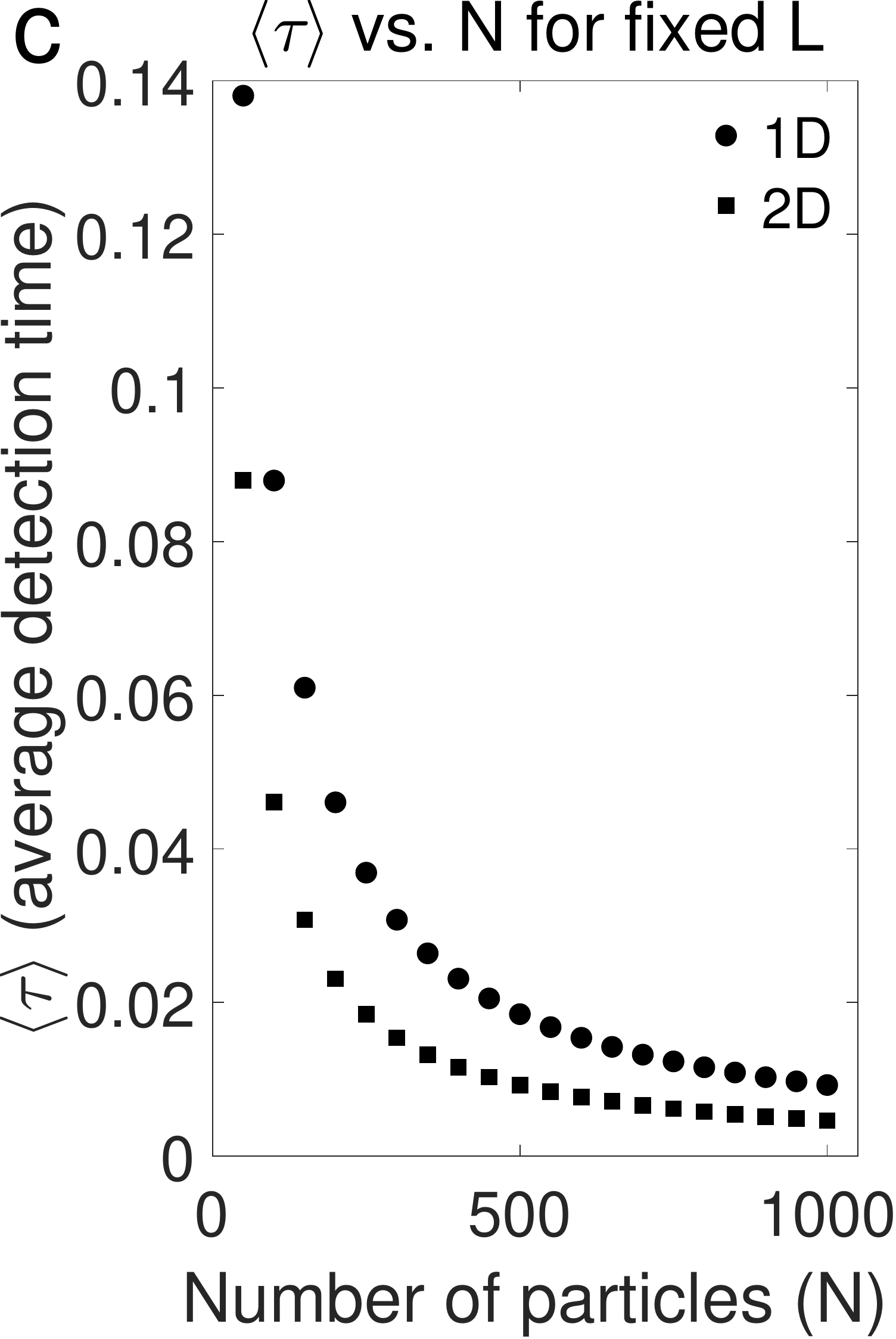}
    \end{subfigure}
    \hspace{.03in}
    \begin{subfigure}[b]{0.31\textwidth}
    \includegraphics[width=\linewidth]{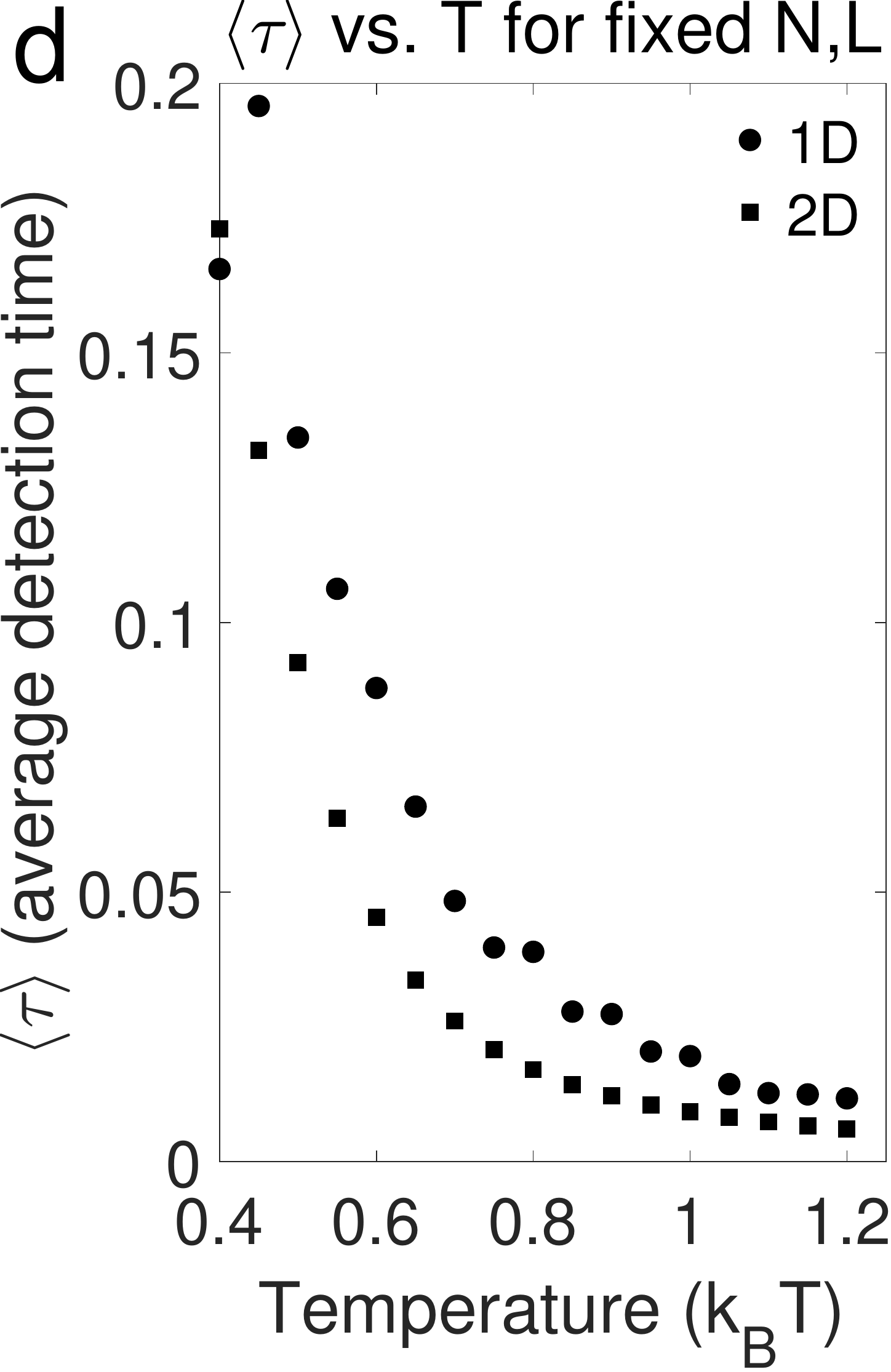}
    \end{subfigure}
    \hspace{.03in}
    \begin{subfigure}[b]{0.33\textwidth}
    \includegraphics[width=\linewidth]{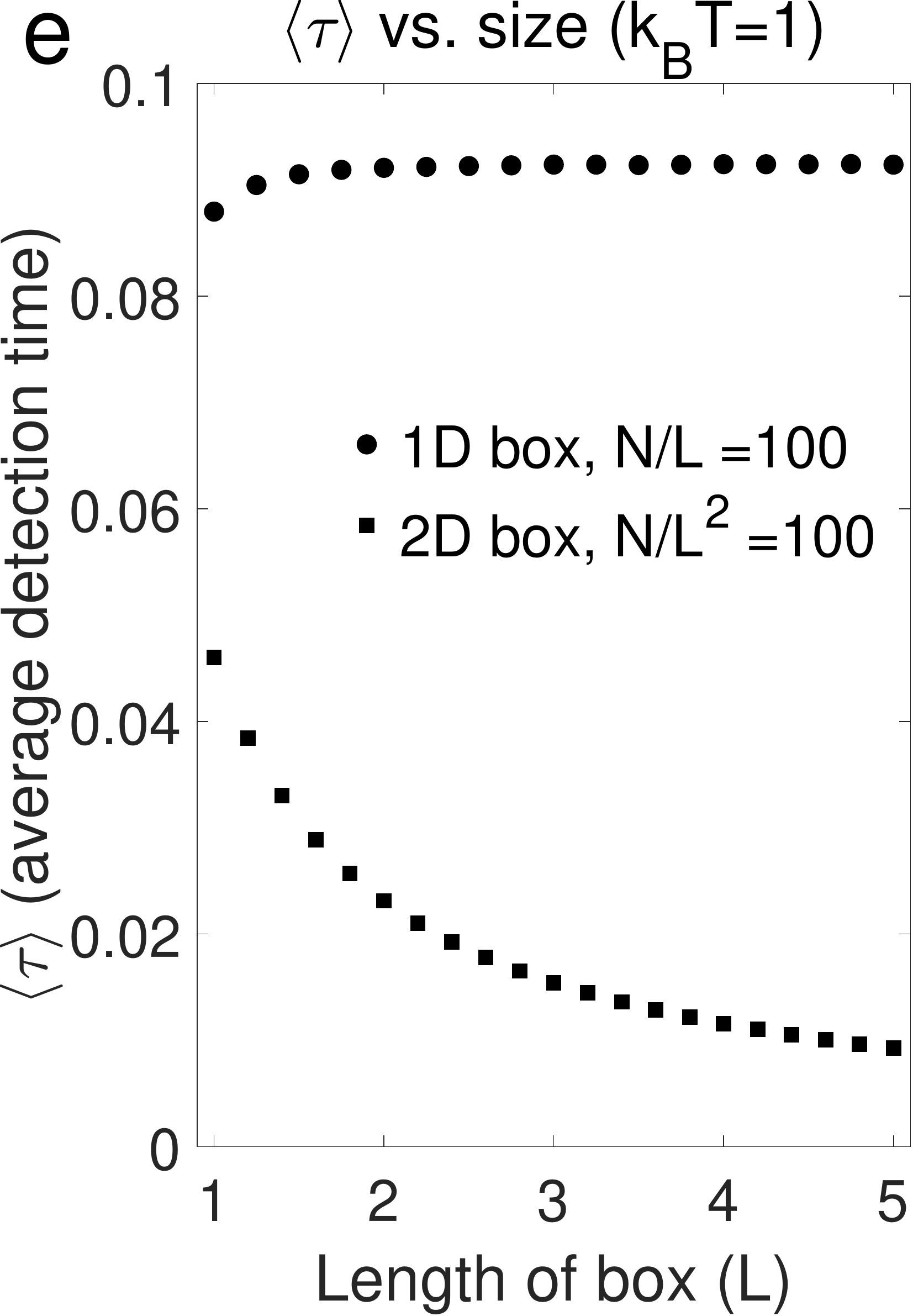}
    \end{subfigure}
    \caption{Probability of detection vs. time ($\tau$) for particles in a 2D fragile container. Panel (a) shows the dependence on temperature for $N=500, L=1$, and the arrow indicates the rare events that are considered in Figs. (\ref{fig:localfig2d}) and (\ref{fig:freedifffig}). Panel (b) shows the dependence on the number of particles, $N$, for $T=1, L=1$. Panels (c), (d) and (e) show the dependence of $\langle \tau \rangle$: (c) on the density of particles, i.e., varying the number of particles in a box of unit size ($T=1, L=1$); (d) temperature ($N=500, L=1$); and (e) on the system size, i.e., the number of particles, while keeping the particle density constant ($T=1$). The bin size of the probability density is $5 \times 10^{-4}$, $v_c=2$, and the units are determined by seting $k_B=m=1$.}
    \label{fig:dettime2d}
\end{figure}

In the one dimensional system, we observe a higher energy density near the detecting end-points for small $\tau$, indicating that high energy particles are close to the end-points and moving towards them. As $\tau$ increases, the high energy particles must be closer to the center, hence a single peak appears near center. For $\tau>(L/2v_c)$ (where $L/2v_c=0.25$ in Fig.\ref{fig:local1d}), the time it takes for a critical velocity particle to go from the center to any one end, we see that two high energy peaks appear. This is because these values of $\tau$ can only arise when all the high velocity particles are close to one of end-points and are moving in a direction opposite to the closest end-point. In 2D, we only observe events where $\tau$ is much smaller than $(L/2v_c)$ since they are already very rare (see Fig.\ref{fig:dettime2d} for the distribution of detection times). In this regime we infer a local mean energy distribution consisting of a combination of the high and low energy peaks similar to the $\tau=0.1$ curve in Fig.\ref{fig:local1d}. Videos of the time dependent energy distribution in 1D and 2D are included as supplementary material. 

Fig.\ref{fig:dettime2d} (a) shows how non-detection events get rarer as $\tau$ increases. We see that non-detection at greater $\tau$ is less likely if the initial temperature of the particles is higher, as we would intuitively expect.

We have explored how the particle number, temperature, and particle density affect the frequency of non-detection events in Fig. \ref{fig:dettime2d}. We observe that the probability density function (PDF) of non-detection of particles is approximately an exponential distribution. We can therefore characterize it by the average detection time, which determines the approximate slope of the lines in Fig. \ref{fig:dettime2d} (a,b). We plot the average detection time as a function of temperature, particle density and system size in Fig. \ref{fig:dettime2d} (c-e). For the 2D ideal gas, as we increase the number of particles (but keeping the density constant) the non-detection events at larger $\tau$ become rarer. In the 2D analog of Eq. \ref{ptau}, as $N \to \infty$, $P_N(\tau) \to \delta (\tau).$

%---------FIGURE 3-------------%
%---------FIGURE 3-------------%

However, in 1D, we see that the effect is reversed. Non-detection for large $\tau$ becomes more common as the system size increases. This difference arises due to the fact that in 2D, the detection region is proportional to the perimeter of the box $(4L)$, while in 1D, it is constant. As the size of the container increases, the number of particles arbitrarily close to the detector increases linearly in 2D, but remains constant in 1D. Therefore, larger non-detection times become less likely in 2D, but not in 1D. In both 1D and 2D, the average detection time decreases as we increase the particle density or the temperature.

Since we are considering non-interacting particles, we can exactly determine the past and future trajectories of all particles given their present state. This allows us to evaluate the probability distribution of the local mean energy at all times from the distribution at any one point of time. This means that we can infer the initial state of a system in which we know that no detection occurred, and predict its future. Fig.\ref{fig:localfig2d} also shows the retrodicted and predicted local mean energy. We see that even if the container walls were to lose their potential to respond to high energy particles at some time, the local temperature variations would persist afterwards.

\section{Corrections to the heat capacity} 

As the particles bounce off the container walls, their direction of motion changes, but their speed remains constant. Since the collisions are elastic, only the component of velocity normal to the wall changes its sign, thus preserving the magnitude of all components of the velocity. This means that a high velocity particle, if any, must already be present in the initial state of the system, thus putting an upper limit on $\tau$. The maximum time required for the first detection, if any, is $\tau_\mathrm{max}=L/v_c$, which occurs when the high velocity particle starts at one edge of the box and is detected at the opposite edge. If no particles are detected before $\tau_\mathrm{max}$, one can infer that there exist no particles with velocity greater than $v_c$ anywhere in the box. This simple inference will have significant thermodynamic outcomes.

For an ideal gas with total energy $E$, the global temperature is given by
\begin{equation}
    T=(\partial S/\partial E)^{-1}=(k_B\partial \log(\Omega)/\partial E)^{-1}.
\end{equation}
where the initial volume of allowed microstates is 
\begin{equation}
    \Omega(t=0) = \frac{L^{3N}}{\hbar^{3N}N!} \!\!\int \mathrm{d}^3\!\vec{v_1} ... \mathrm{d}^3 \vec{v_N} \delta [(m/2)(\vec{v_1}^2 + ... + \vec{v_N}^2) - E].
\end{equation}

This integral is the surface area of a $3N$ dimensional sphere. For $t>\tau_\mathrm{max}$, since there can be no velocity component greater than $v_c$, the volume of allowed microstates becomes
\begin{equation}\label{spherecubeint}
    \begin{split}
    \Omega(t) = \frac{L^{3N}}{\hbar^{3N}N!} \!\!\int  \!\Big[ &\delta \{(m/2)(\vec{v_1}^2 + ... + \vec{v_N}^2) - E\}
     \theta (\vec{v_1}-\vec{v_c}) ... \theta(\vec{v_N}-\vec{v_c}) \Big] \mathrm{d}^3 \vec{v_1} ... \mathrm{d}^3 \vec{v_N}.
    \end{split}
\end{equation}

%---------FIGURE 4-------------%
\begin{figure}
    \centering
    \includegraphics[width=0.7\linewidth]{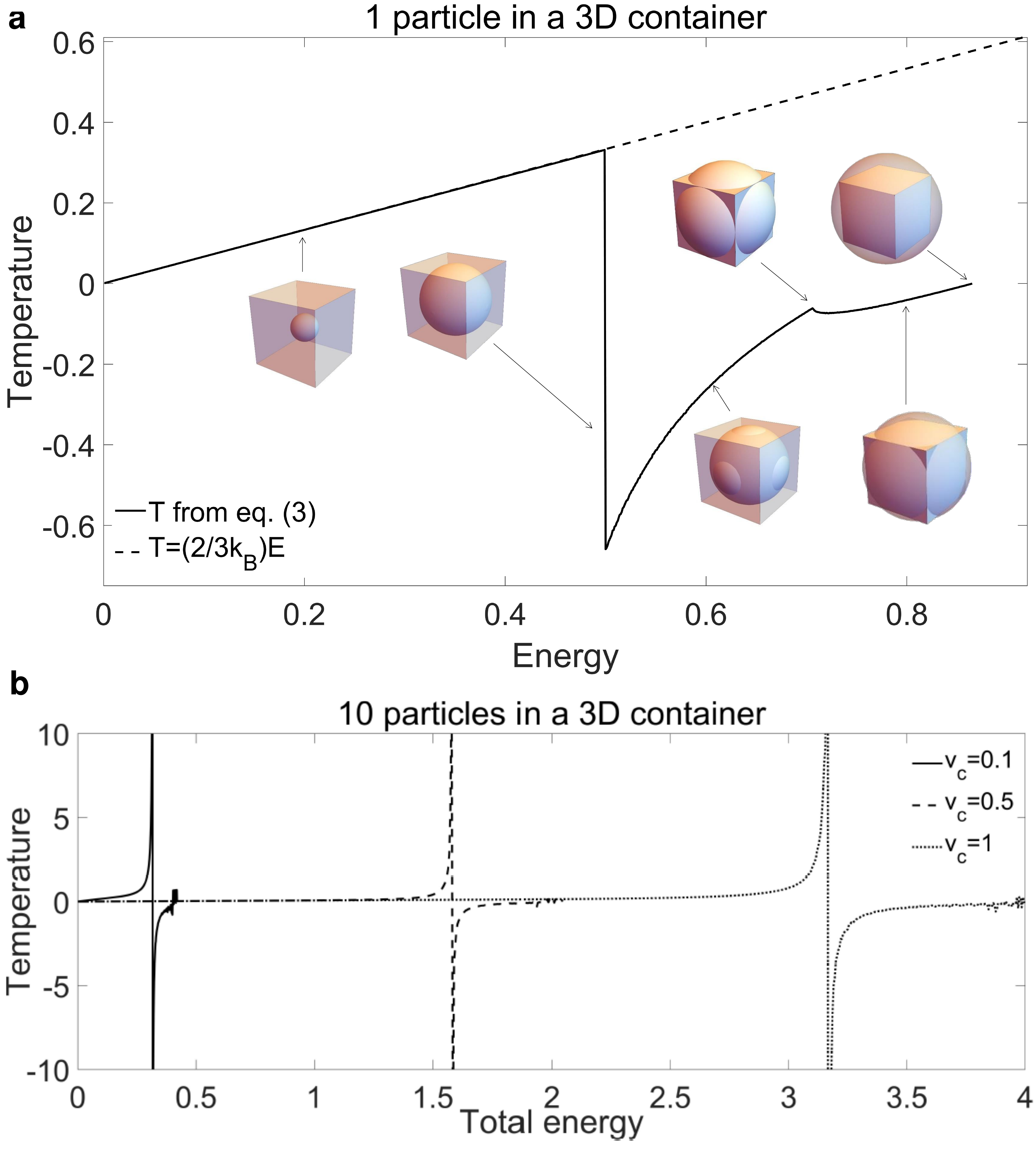}
    \caption{(a) Comparison between the temperature-energy relations for an ideal gas in a container with (dashed line) and without (solid line) fragile walls ($v_c=2$). We geometrically illustrate why there are singular points in the energy-temperature curve in terms of the phase space of the system of a single particle. (b) Temperature-energy relations for 10 particles in a fragile container for various $v_c$. $k_B=L=m=1$, and the energy, $E$, is measured in units of $E/2E_c$.}
    \label{fig:globalfig}
\end{figure}
%---------FIGURE 4-------------%
This integral is the surface area of intersection of a $3N$ dimensional sphere with a $3N$ dimensional cube. The radius of the sphere depends on the total energy and the side of the cube depends on the critical velocity. Analytical formulas for this surface area can be found in \cite{xu1996volume}. 

Changing the total energy ($E$) of the particles inside the box corresponds to changing the radius of the $3N-$sphere. The volume of allowed microstates, $\Omega (E)$, is not differentiable at points where the sphere touches various hyperplanes of intersection of the faces of the cube. This causes the temperature to be a discontinuous function of energy, as shown in Fig.\ref{fig:globalfig}. For example, for the special case of a single particle (where the velocity space is three dimensional), these singularities occur when the sphere touches the faces, edges and vertices of the cube (see Fig.\ref{fig:globalfig}(a)).

At higher dimensions (i.e. for systems with larger number of particles) there will be a larger number of such singularities, as we move up the energy scale, i.e. whenever the velocity hypersphere crosses hyper-faces, lesser dimensional hyper-faces, faces, edges and finally the corners of the hypercube.
The heat capacity $C_v$, which is the derivative of energy with respect to temperature, has singularities at these points. 

When the total energy is very small, we do not see any deviation from the ideal gas law. This is because even if all the energy were due to one velocity component of a single particle, the velocity would not exceed $v_c$. When the total energy exceeds a critical value, the temperature of the system becomes negative. This indicates that the system is in a population-inverted state and confirms that it must be out of equilibrium \cite{swendsen2015gibbs,frenkel2015gibbs,cerino2015consistent}. This transition occurs when the high total energy along with the constraint, $v<v_c$ for every particle causes there to be more high energy particles than low energy particles. Finally, when the energy is so large that there can be no particles with $v<v_c$, the number of microstates becomes exactly zero, implying that such an event is impossible.

We evaluate the integral in Eq.\ref{spherecubeint} by Monte Carlo integration by generating $10^9$ sets of random $3N-$dimensional vectors with norm $\textstyle{\sqrt{2m/E}}$ and finding the fraction of sets where no component of the vector exceeds $v_c$. We then multiply this by the surface area of the entire sphere to get the area of intersection. 

%---------FIGURE 5-------------%
\begin{figure}
%    \centering
    \includegraphics[width=\linewidth]{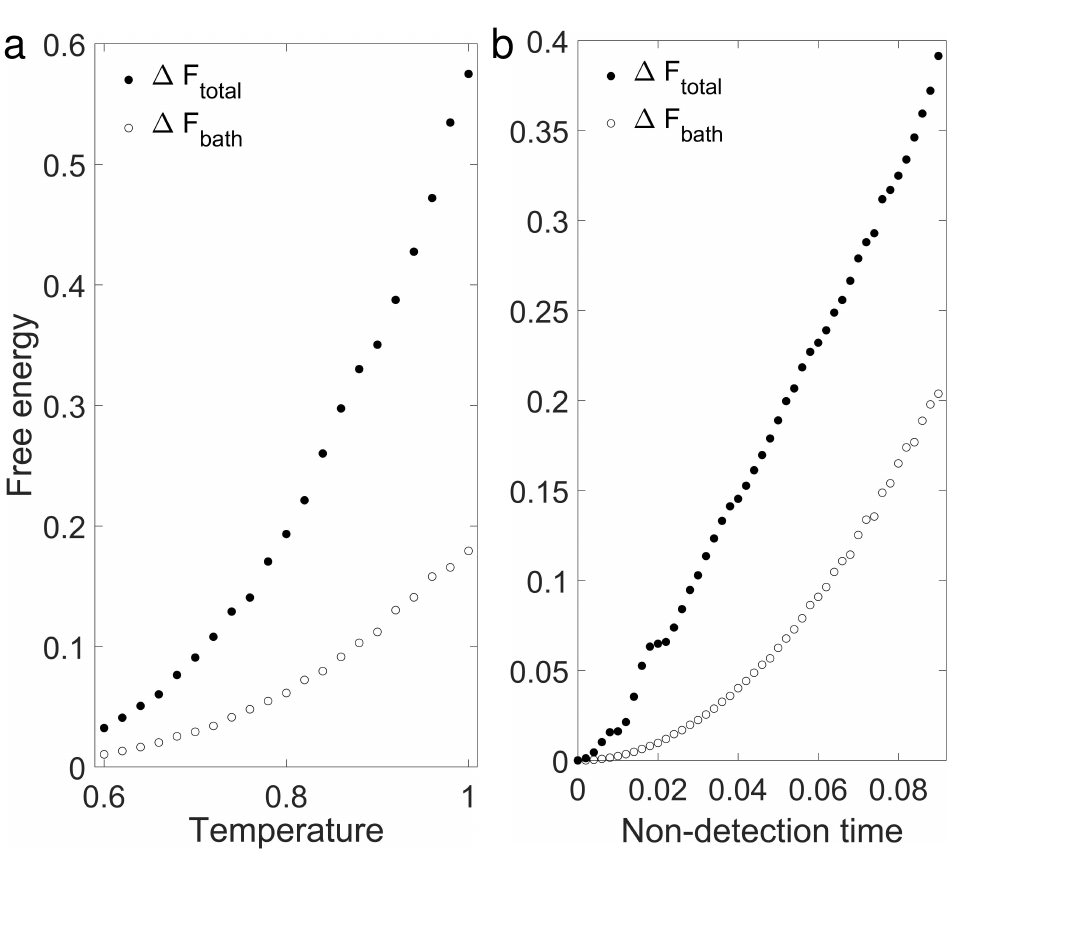}
    \caption{Free energy of a 2D gas in a fragile container. (a) Temperature dependence of free energy at $\tau=0.05$. (b) Variation of free energy with $\tau$ for $k_BT=0.8$. Each data point was obtained by $5\times 10^6$ Monte Carlo samples. $v_c=2$, $L=1$, $N=500$.  $k_B=l=m=1$.}
    \label{fig:freedifffig}
\end{figure}
%---------FIGURE 5-------------%
\section{Work extraction}
Local differences in energy, such as those shown in Figs. \ref{fig:local1d} and \ref{fig:localfig2d} can be used to extract work. For example, after some non-detection time $\tau$ one may remove the detecting parts from the walls, and place turbines or balloons between the inferred warmer and colder regions. The turbines will turn or the balloons will expand as the container thermalizes back to a uniform energy distribution. Subsequently, we can extract additional work if the container as a whole is cooler or warmer than the heat bath from which it was sampled. In this paper, we will not concern ourselves with practical techniques to extract work, as they have already been studied \cite{koski2015chip,aaberg2013truly,procaccia1976potential,sagawa2009minimal}. 

We now proceed to finding the maximum possible work that can be extracted from the system using the inferences derived from the non-event. The maximum work that can be extracted from the system by coupling it to the heat bath from which the particles were sampled is equal to the free energy difference between the inferred state and the equilibrium state.

The maximum work that can be extracted from a non-equilibrium state of a system at a fixed temperature is equal to the Helmholtz free energy. The simple expression for free energy in terms of the average energy and entropy allows us to find the maximum work that can be extracted from the system without the need for specific details about the work extraction mechanism. We use the Sackur-Tetrode equation to find the local entropy of the ideal gas. Since the particles under consideration are non-interacting, the energy and entropy are additive. Therefore, the total energy and total entropy are the integral of the local energy and entropy densities. Using the standard definition of free energy, and the equation for the entropy of an ideal gas in $d-$dimensions,
\begin{equation}\label{freediff}
    \Delta F = \Delta U - T \Delta S = \int \mathrm{d}^{\textcolor{blue}{d}}x \bigg[(U(x)-\Bar{U})-\Bar{U}\log{\Big(\frac{U(x)}{\bar{U}}\Big)}\bigg].
\end{equation}
Here, $\Bar{U}=(d/2L^d)Nk_BT$ is the \emph{a-priori} mean energy density and $U(x)$ is the inferred mean energy density at $x$ (plotted in Figs. \ref{fig:local1d} and \ref{fig:localfig2d}). We have also assumed constant particle density in space.

Fig. \ref{fig:freedifffig} shows the variation of $\Delta F$ with bath temperature $T$ and non-detection time $\tau$. We observe that as inferred state becomes rarer, we can extract more work from it. $\Delta F_{bath}$ is the amount of work that can be extracted only due to the container as a whole having a different temperature than the bath its particles were sampled from.
As $\tau\to 0$ at a fixed temperature, non-detection is more common and system has not yet moved far from equilibrium. Therefore the free energy differences decrease to zero. When $\tau=0$, we have no information about the system, hence the free energy differences are exactly zero.

\section{Discussion} 
Any finite ensemble is expected to fluctuate about its equilibrium state \cite{hickman2016temperature,mishin2015thermodynamic,bertini2015macroscopic,qian2001relative}. Then the non-detection of macroscopic changes in the finite system informs us of the degree of the departure from equilibrium as much as the detection of such changes.

As the number of particles increases, the probability of non-detection events decreases. Therefore the effects considered in Section 4 can be more easily observed in systems with a small number of particles. Many other finite size effects such as second law violations also vanish in the thermodynamic limit \cite{jarzynski1997equilibrium,jarzynski1997nonequilibrium,crooks1999entropy,kurchan1998fluctuation,seifert2005entropy,evans1993probability,evans1994equilibrium,evans2002fluctuation,gallavotti1995dynamical}.

Extracting work from thermodynamic fluctuations, as shown in section 3, is not by virtue of finite size/number but rather, by virtue of finite number density. In systems where the detector size does not scale with the particle number, such as in case of a single explosive particle inside an ideal gas, the deviations from equilibrium will occur even if the system size approaches infinity with a constant particle density. We also see this in Fig. \ref{fig:dettime2d} (e), where $\tau$ increases and then becomes constant in 1D as the system size increases.

In this article, we only considered non-interacting particles, which played a key role in the results.
% discussed in Section 4.
If the particles interact, then there is no upper limit on the time of the first detection. Even if the initial configuration, as drawn from the bath, does not contain any high velocity particles, the interactions can produce one at any time. Conversely, a high velocity particle initially present in the system is likely to dissipate its energy before being detected by the walls. Therefore, the local and global thermodynamic effects of non-detection will be different,  and this remains an open problem.

As we see in Fig.\ref{fig:localfig2d}, the local energy distribution has features that depend on the shape of the container. This effect may also be extended to quantum mechanics, where confinement is already known to play a role. In particular, the effect of finite size and confinement geometry on quantum gases and their phase transitions has been studied \cite{pathria1983phase,subrahmanyam1989finite,dey2020finite,bhattacharyya1998finite,danchev2000theory}.

Here we used the Bayes theorem to find the local mean energy of the system. Bayesian methods and the general relation between thermodynamics and information theory were developed in \cite{jaynes1957information,jaynes1957information2,jaynes1968prior,jaynes1988relation}. However, the use of Bayesian methods to determine thermodynamic quantities can lead to conceptual problems \cite{shalizi2004backwards}.

While we only considered the local mean energy in the article, the problem of using it to define a local temperature has been a subject of considerable debate. Conventionally, local temperature is defined either through a mapping to the ensemble average of energy, or as the temperature measured by a model thermometer after coupling for a sufficiently long time. For further discussion on this issue, see \cite{casas2003temperature,brites2012thermometry,hartmann2004local,hartmann2004existence,hartmann2005measurable,ghonge2018temperature}.

\textbf{Declaration of competing interest}

The authors declare that they have no known competing financial interests or personal relationships that could have appeared to influence the work reported in this paper.

%\bibliography{references}

\begin{thebibliography}{70}%
\makeatletter
\providecommand \@ifxundefined [1]{%
 \@ifx{#1\undefined}
}%
\providecommand \@ifnum [1]{%
 \ifnum #1\expandafter \@firstoftwo
 \else \expandafter \@secondoftwo
 \fi
}%
\providecommand \@ifx [1]{%
 \ifx #1\expandafter \@firstoftwo
 \else \expandafter \@secondoftwo
 \fi
}%
\providecommand \natexlab [1]{#1}%
\providecommand \enquote  [1]{``#1''}%
\providecommand \bibnamefont  [1]{#1}%
\providecommand \bibfnamefont [1]{#1}%
\providecommand \citenamefont [1]{#1}%
\providecommand \href@noop [0]{\@secondoftwo}%
\providecommand \href [0]{\begingroup \@sanitize@url \@href}%
\providecommand \@href[1]{\@@startlink{#1}\@@href}%
\providecommand \@@href[1]{\endgroup#1\@@endlink}%
\providecommand \@sanitize@url [0]{\catcode `\\12\catcode `\$12\catcode
  `\&12\catcode `\#12\catcode `\^12\catcode `\_12\catcode `\%12\relax}%
\providecommand \@@startlink[1]{}%
\providecommand \@@endlink[0]{}%
\providecommand \url  [0]{\begingroup\@sanitize@url \@url }%
\providecommand \@url [1]{\endgroup\@href {#1}{\urlprefix }}%
\providecommand \urlprefix  [0]{URL }%
\providecommand \Eprint [0]{\href }%
\providecommand \doibase [0]{http://dx.doi.org/}%
\providecommand \selectlanguage [0]{\@gobble}%
\providecommand \bibinfo  [0]{\@secondoftwo}%
\providecommand \bibfield  [0]{\@secondoftwo}%
\providecommand \translation [1]{[#1]}%
\providecommand \BibitemOpen [0]{}%
\providecommand \bibitemStop [0]{}%
\providecommand \bibitemNoStop [0]{.\EOS\space}%
\providecommand \EOS [0]{\spacefactor3000\relax}%
\providecommand \BibitemShut  [1]{\csname bibitem#1\endcsname}%
\let\auto@bib@innerbib\@empty
%</preamble>
\bibitem [{\citenamefont {Reiss}(2012)}]{reiss2012methods}%
  \BibitemOpen
  \bibfield  {author} {\bibinfo {author} {\bibfnamefont {H.}~\bibnamefont
  {Reiss}},\ }\href@noop {} {\emph {\bibinfo {title} {Methods of
  thermodynamics}}}\ (\bibinfo  {publisher} {Courier Corporation},\ \bibinfo
  {year} {2012})\BibitemShut {NoStop}%
\bibitem [{\citenamefont {Pitowsky}(2006)}]{pitowsky2006definition}%
  \BibitemOpen
  \bibfield  {author} {\bibinfo {author} {\bibfnamefont {I.}~\bibnamefont
  {Pitowsky}},\ }\href@noop {} {\bibfield  {journal} {\bibinfo  {journal}
  {Studies in History and Philosophy of Science Part B: Studies in History and
  Philosophy of Modern Physics}\ }\textbf {\bibinfo {volume} {37}},\ \bibinfo
  {pages} {431} (\bibinfo {year} {2006})}\BibitemShut {NoStop}%
\bibitem [{\citenamefont {Werndl}\ and\ \citenamefont
  {Frigg}(2015)}]{werndl2015rethinking}%
  \BibitemOpen
  \bibfield  {author} {\bibinfo {author} {\bibfnamefont {C.}~\bibnamefont
  {Werndl}}\ and\ \bibinfo {author} {\bibfnamefont {R.}~\bibnamefont {Frigg}},\
  }\href@noop {} {\bibfield  {journal} {\bibinfo  {journal} {Philosophy of
  Science}\ }\textbf {\bibinfo {volume} {82}},\ \bibinfo {pages} {1224}
  (\bibinfo {year} {2015})}\BibitemShut {NoStop}%
\bibitem [{\citenamefont {Puglisi}\ and\ \citenamefont {Marini
  Bettolo~Marconi}(2017)}]{puglisi2017clausius}%
  \BibitemOpen
  \bibfield  {author} {\bibinfo {author} {\bibfnamefont {A.}~\bibnamefont
  {Puglisi}}\ and\ \bibinfo {author} {\bibfnamefont {U.}~\bibnamefont {Marini
  Bettolo~Marconi}},\ }\href@noop {} {\bibfield  {journal} {\bibinfo  {journal}
  {Entropy}\ }\textbf {\bibinfo {volume} {19}},\ \bibinfo {pages} {356}
  (\bibinfo {year} {2017})}\BibitemShut {NoStop}%
\bibitem [{\citenamefont {Puglisi}\ \emph {et~al.}(2017)\citenamefont
  {Puglisi}, \citenamefont {Sarracino},\ and\ \citenamefont
  {Vulpiani}}]{puglisi2017temperature}%
  \BibitemOpen
  \bibfield  {author} {\bibinfo {author} {\bibfnamefont {A.}~\bibnamefont
  {Puglisi}}, \bibinfo {author} {\bibfnamefont {A.}~\bibnamefont {Sarracino}},
  \ and\ \bibinfo {author} {\bibfnamefont {A.}~\bibnamefont {Vulpiani}},\
  }\href@noop {} {\bibfield  {journal} {\bibinfo  {journal} {Physics Reports}\
  }\textbf {\bibinfo {volume} {709-710}},\ \bibinfo {pages} {1} (\bibinfo
  {year} {2017})}\BibitemShut {NoStop}%
\bibitem [{\citenamefont {Rupprecht}\ and\ \citenamefont
  {Vural}(2019{\natexlab{a}})}]{rupprecht2019maxwell}%
  \BibitemOpen
  \bibfield  {author} {\bibinfo {author} {\bibfnamefont {N.}~\bibnamefont
  {Rupprecht}}\ and\ \bibinfo {author} {\bibfnamefont {D.~C.}\ \bibnamefont
  {Vural}},\ }\href {\doibase 10.1103/PhysRevLett.123.080603} {\bibfield
  {journal} {\bibinfo  {journal} {Phys. Rev. Lett.}\ }\textbf {\bibinfo
  {volume} {123}},\ \bibinfo {pages} {080603} (\bibinfo {year}
  {2019}{\natexlab{a}})}\BibitemShut {NoStop}%
\bibitem [{\citenamefont {Koski}\ \emph {et~al.}(2015)\citenamefont {Koski},
  \citenamefont {Kutvonen}, \citenamefont {Khaymovich}, \citenamefont
  {Ala-Nissila},\ and\ \citenamefont {Pekola}}]{koski2015chip}%
  \BibitemOpen
  \bibfield  {author} {\bibinfo {author} {\bibfnamefont {J.~V.}\ \bibnamefont
  {Koski}}, \bibinfo {author} {\bibfnamefont {A.}~\bibnamefont {Kutvonen}},
  \bibinfo {author} {\bibfnamefont {I.~M.}\ \bibnamefont {Khaymovich}},
  \bibinfo {author} {\bibfnamefont {T.}~\bibnamefont {Ala-Nissila}}, \ and\
  \bibinfo {author} {\bibfnamefont {J.~P.}\ \bibnamefont {Pekola}},\
  }\href@noop {} {\bibfield  {journal} {\bibinfo  {journal} {Physical review
  letters}\ }\textbf {\bibinfo {volume} {115}},\ \bibinfo {pages} {260602}
  (\bibinfo {year} {2015})}\BibitemShut {NoStop}%
\bibitem [{\citenamefont {{\AA}berg}(2013)}]{aaberg2013truly}%
  \BibitemOpen
  \bibfield  {author} {\bibinfo {author} {\bibfnamefont {J.}~\bibnamefont
  {{\AA}berg}},\ }\href@noop {} {\bibfield  {journal} {\bibinfo  {journal}
  {Nature communications}\ }\textbf {\bibinfo {volume} {4}},\ \bibinfo {pages}
  {1925} (\bibinfo {year} {2013})}\BibitemShut {NoStop}%
\bibitem [{\citenamefont {Parrondo}\ \emph {et~al.}(2015)\citenamefont
  {Parrondo}, \citenamefont {Horowitz},\ and\ \citenamefont
  {Sagawa}}]{parrondo2015thermodynamics}%
  \BibitemOpen
  \bibfield  {author} {\bibinfo {author} {\bibfnamefont {J.~M.}\ \bibnamefont
  {Parrondo}}, \bibinfo {author} {\bibfnamefont {J.~M.}\ \bibnamefont
  {Horowitz}}, \ and\ \bibinfo {author} {\bibfnamefont {T.}~\bibnamefont
  {Sagawa}},\ }\href@noop {} {\bibfield  {journal} {\bibinfo  {journal} {Nature
  physics}\ }\textbf {\bibinfo {volume} {11}},\ \bibinfo {pages} {131}
  (\bibinfo {year} {2015})}\BibitemShut {NoStop}%
\bibitem [{\citenamefont {Rupprecht}\ and\ \citenamefont
  {Vural}(2018)}]{rupprecht2018limits}%
  \BibitemOpen
  \bibfield  {author} {\bibinfo {author} {\bibfnamefont {N.}~\bibnamefont
  {Rupprecht}}\ and\ \bibinfo {author} {\bibfnamefont {D.~C.}\ \bibnamefont
  {Vural}},\ }\href@noop {} {\bibfield  {journal} {\bibinfo  {journal}
  {Physical Review E}\ }\textbf {\bibinfo {volume} {97}},\ \bibinfo {pages}
  {062155} (\bibinfo {year} {2018})}\BibitemShut {NoStop}%
\bibitem [{\citenamefont {Rupprecht}\ and\ \citenamefont
  {Vural}(2019{\natexlab{b}})}]{rupprecht2019enhancing}%
  \BibitemOpen
  \bibfield  {author} {\bibinfo {author} {\bibfnamefont {N.}~\bibnamefont
  {Rupprecht}}\ and\ \bibinfo {author} {\bibfnamefont {D.~C.}\ \bibnamefont
  {Vural}},\ }\href@noop {} {\bibfield  {journal} {\bibinfo  {journal}
  {Communications Physics}\ }\textbf {\bibinfo {volume} {2}},\ \bibinfo {pages}
  {57} (\bibinfo {year} {2019}{\natexlab{b}})}\BibitemShut {NoStop}%
\bibitem [{\citenamefont {Hartmann}\ \emph
  {et~al.}(2004{\natexlab{a}})\citenamefont {Hartmann}, \citenamefont
  {Mahler},\ and\ \citenamefont {Hess}}]{hartmann2004local}%
  \BibitemOpen
  \bibfield  {author} {\bibinfo {author} {\bibfnamefont {M.}~\bibnamefont
  {Hartmann}}, \bibinfo {author} {\bibfnamefont {G.}~\bibnamefont {Mahler}}, \
  and\ \bibinfo {author} {\bibfnamefont {O.}~\bibnamefont {Hess}},\ }\href@noop
  {} {\bibfield  {journal} {\bibinfo  {journal} {Physical Review E}\ }\textbf
  {\bibinfo {volume} {70}},\ \bibinfo {pages} {066148} (\bibinfo {year}
  {2004}{\natexlab{a}})}\BibitemShut {NoStop}%
\bibitem [{\citenamefont {Brites}\ \emph {et~al.}(2012)\citenamefont {Brites},
  \citenamefont {Lima}, \citenamefont {Silva}, \citenamefont {Mill{\'a}n},
  \citenamefont {Amaral}, \citenamefont {Palacio},\ and\ \citenamefont
  {Carlos}}]{brites2012thermometry}%
  \BibitemOpen
  \bibfield  {author} {\bibinfo {author} {\bibfnamefont {C.~D.}\ \bibnamefont
  {Brites}}, \bibinfo {author} {\bibfnamefont {P.~P.}\ \bibnamefont {Lima}},
  \bibinfo {author} {\bibfnamefont {N.~J.}\ \bibnamefont {Silva}}, \bibinfo
  {author} {\bibfnamefont {A.}~\bibnamefont {Mill{\'a}n}}, \bibinfo {author}
  {\bibfnamefont {V.~S.}\ \bibnamefont {Amaral}}, \bibinfo {author}
  {\bibfnamefont {F.}~\bibnamefont {Palacio}}, \ and\ \bibinfo {author}
  {\bibfnamefont {L.~D.}\ \bibnamefont {Carlos}},\ }\href@noop {} {\bibfield
  {journal} {\bibinfo  {journal} {Nanoscale}\ }\textbf {\bibinfo {volume}
  {4}},\ \bibinfo {pages} {4799} (\bibinfo {year} {2012})}\BibitemShut
  {NoStop}%
\bibitem [{\citenamefont {M{\"u}ller}(2019)}]{muller2019information}%
  \BibitemOpen
  \bibfield  {author} {\bibinfo {author} {\bibfnamefont {J.~G.}\ \bibnamefont
  {M{\"u}ller}},\ }\href@noop {} {\bibfield  {journal} {\bibinfo  {journal}
  {Entropy}\ }\textbf {\bibinfo {volume} {21}},\ \bibinfo {pages} {1052}
  (\bibinfo {year} {2019})}\BibitemShut {NoStop}%
\bibitem [{\citenamefont {Naghiloo}\ \emph {et~al.}(2018)\citenamefont
  {Naghiloo}, \citenamefont {Alonso}, \citenamefont {Romito}, \citenamefont
  {Lutz},\ and\ \citenamefont {Murch}}]{naghiloo2018information}%
  \BibitemOpen
  \bibfield  {author} {\bibinfo {author} {\bibfnamefont {M.}~\bibnamefont
  {Naghiloo}}, \bibinfo {author} {\bibfnamefont {J.}~\bibnamefont {Alonso}},
  \bibinfo {author} {\bibfnamefont {A.}~\bibnamefont {Romito}}, \bibinfo
  {author} {\bibfnamefont {E.}~\bibnamefont {Lutz}}, \ and\ \bibinfo {author}
  {\bibfnamefont {K.}~\bibnamefont {Murch}},\ }\href@noop {} {\bibfield
  {journal} {\bibinfo  {journal} {Physical review letters}\ }\textbf {\bibinfo
  {volume} {121}},\ \bibinfo {pages} {030604} (\bibinfo {year}
  {2018})}\BibitemShut {NoStop}%
\bibitem [{\citenamefont {Hickman}\ and\ \citenamefont
  {Mishin}(2016)}]{hickman2016temperature}%
  \BibitemOpen
  \bibfield  {author} {\bibinfo {author} {\bibfnamefont {J.}~\bibnamefont
  {Hickman}}\ and\ \bibinfo {author} {\bibfnamefont {Y.}~\bibnamefont
  {Mishin}},\ }\href@noop {} {\bibfield  {journal} {\bibinfo  {journal}
  {Physical Review B}\ }\textbf {\bibinfo {volume} {94}},\ \bibinfo {pages}
  {184311} (\bibinfo {year} {2016})}\BibitemShut {NoStop}%
\bibitem [{\citenamefont {Mishin}(2015)}]{mishin2015thermodynamic}%
  \BibitemOpen
  \bibfield  {author} {\bibinfo {author} {\bibfnamefont {Y.}~\bibnamefont
  {Mishin}},\ }\href@noop {} {\bibfield  {journal} {\bibinfo  {journal} {Annals
  of Physics}\ }\textbf {\bibinfo {volume} {363}},\ \bibinfo {pages} {48}
  (\bibinfo {year} {2015})}\BibitemShut {NoStop}%
\bibitem [{\citenamefont {Bertini}\ \emph {et~al.}(2015)\citenamefont
  {Bertini}, \citenamefont {De~Sole}, \citenamefont {Gabrielli}, \citenamefont
  {Jona-Lasinio},\ and\ \citenamefont {Landim}}]{bertini2015macroscopic}%
  \BibitemOpen
  \bibfield  {author} {\bibinfo {author} {\bibfnamefont {L.}~\bibnamefont
  {Bertini}}, \bibinfo {author} {\bibfnamefont {A.}~\bibnamefont {De~Sole}},
  \bibinfo {author} {\bibfnamefont {D.}~\bibnamefont {Gabrielli}}, \bibinfo
  {author} {\bibfnamefont {G.}~\bibnamefont {Jona-Lasinio}}, \ and\ \bibinfo
  {author} {\bibfnamefont {C.}~\bibnamefont {Landim}},\ }\href@noop {}
  {\bibfield  {journal} {\bibinfo  {journal} {Reviews of Modern Physics}\
  }\textbf {\bibinfo {volume} {87}},\ \bibinfo {pages} {593} (\bibinfo {year}
  {2015})}\BibitemShut {NoStop}%
\bibitem [{\citenamefont {Qian}(2001)}]{qian2001relative}%
  \BibitemOpen
  \bibfield  {author} {\bibinfo {author} {\bibfnamefont {H.}~\bibnamefont
  {Qian}},\ }\href@noop {} {\bibfield  {journal} {\bibinfo  {journal} {Physical
  Review E}\ }\textbf {\bibinfo {volume} {63}},\ \bibinfo {pages} {042103}
  (\bibinfo {year} {2001})}\BibitemShut {NoStop}%
\bibitem [{\citenamefont
  {Jarzynski}(1997{\natexlab{a}})}]{jarzynski1997equilibrium}%
  \BibitemOpen
  \bibfield  {author} {\bibinfo {author} {\bibfnamefont {C.}~\bibnamefont
  {Jarzynski}},\ }\href@noop {} {\bibfield  {journal} {\bibinfo  {journal}
  {Physical Review E}\ }\textbf {\bibinfo {volume} {56}},\ \bibinfo {pages}
  {5018} (\bibinfo {year} {1997}{\natexlab{a}})}\BibitemShut {NoStop}%
\bibitem [{\citenamefont
  {Jarzynski}(1997{\natexlab{b}})}]{jarzynski1997nonequilibrium}%
  \BibitemOpen
  \bibfield  {author} {\bibinfo {author} {\bibfnamefont {C.}~\bibnamefont
  {Jarzynski}},\ }\href@noop {} {\bibfield  {journal} {\bibinfo  {journal}
  {Physical Review Letters}\ }\textbf {\bibinfo {volume} {78}},\ \bibinfo
  {pages} {2690} (\bibinfo {year} {1997}{\natexlab{b}})}\BibitemShut {NoStop}%
\bibitem [{\citenamefont {Crooks}(1999)}]{crooks1999entropy}%
  \BibitemOpen
  \bibfield  {author} {\bibinfo {author} {\bibfnamefont {G.~E.}\ \bibnamefont
  {Crooks}},\ }\href@noop {} {\bibfield  {journal} {\bibinfo  {journal}
  {Physical Review E}\ }\textbf {\bibinfo {volume} {60}},\ \bibinfo {pages}
  {2721} (\bibinfo {year} {1999})}\BibitemShut {NoStop}%
\bibitem [{\citenamefont {Kurchan}(1998)}]{kurchan1998fluctuation}%
  \BibitemOpen
  \bibfield  {author} {\bibinfo {author} {\bibfnamefont {J.}~\bibnamefont
  {Kurchan}},\ }\href@noop {} {\bibfield  {journal} {\bibinfo  {journal}
  {Journal of Physics A: Mathematical and General}\ }\textbf {\bibinfo {volume}
  {31}},\ \bibinfo {pages} {3719} (\bibinfo {year} {1998})}\BibitemShut
  {NoStop}%
\bibitem [{\citenamefont {Seifert}(2005)}]{seifert2005entropy}%
  \BibitemOpen
  \bibfield  {author} {\bibinfo {author} {\bibfnamefont {U.}~\bibnamefont
  {Seifert}},\ }\href@noop {} {\bibfield  {journal} {\bibinfo  {journal}
  {Physical review letters}\ }\textbf {\bibinfo {volume} {95}},\ \bibinfo
  {pages} {040602} (\bibinfo {year} {2005})}\BibitemShut {NoStop}%
\bibitem [{\citenamefont {Evans}\ \emph {et~al.}(1993)\citenamefont {Evans},
  \citenamefont {Cohen},\ and\ \citenamefont {Morriss}}]{evans1993probability}%
  \BibitemOpen
  \bibfield  {author} {\bibinfo {author} {\bibfnamefont {D.~J.}\ \bibnamefont
  {Evans}}, \bibinfo {author} {\bibfnamefont {E.~G.~D.}\ \bibnamefont {Cohen}},
  \ and\ \bibinfo {author} {\bibfnamefont {G.~P.}\ \bibnamefont {Morriss}},\
  }\href@noop {} {\bibfield  {journal} {\bibinfo  {journal} {Physical review
  letters}\ }\textbf {\bibinfo {volume} {71}},\ \bibinfo {pages} {2401}
  (\bibinfo {year} {1993})}\BibitemShut {NoStop}%
\bibitem [{\citenamefont {Wang}\ \emph {et~al.}(2002)\citenamefont {Wang},
  \citenamefont {Sevick}, \citenamefont {Mittag}, \citenamefont {Searles},\
  and\ \citenamefont {Evans}}]{wang2002experimental}%
  \BibitemOpen
  \bibfield  {author} {\bibinfo {author} {\bibfnamefont {G.}~\bibnamefont
  {Wang}}, \bibinfo {author} {\bibfnamefont {E.~M.}\ \bibnamefont {Sevick}},
  \bibinfo {author} {\bibfnamefont {E.}~\bibnamefont {Mittag}}, \bibinfo
  {author} {\bibfnamefont {D.~J.}\ \bibnamefont {Searles}}, \ and\ \bibinfo
  {author} {\bibfnamefont {D.~J.}\ \bibnamefont {Evans}},\ }\href@noop {}
  {\bibfield  {journal} {\bibinfo  {journal} {Physical Review Letters}\
  }\textbf {\bibinfo {volume} {89}},\ \bibinfo {pages} {050601} (\bibinfo
  {year} {2002})}\BibitemShut {NoStop}%
\bibitem [{\citenamefont {Evans}\ and\ \citenamefont
  {Searles}(1994)}]{evans1994equilibrium}%
  \BibitemOpen
  \bibfield  {author} {\bibinfo {author} {\bibfnamefont {D.~J.}\ \bibnamefont
  {Evans}}\ and\ \bibinfo {author} {\bibfnamefont {D.~J.}\ \bibnamefont
  {Searles}},\ }\href@noop {} {\bibfield  {journal} {\bibinfo  {journal}
  {Physical Review E}\ }\textbf {\bibinfo {volume} {50}},\ \bibinfo {pages}
  {1645} (\bibinfo {year} {1994})}\BibitemShut {NoStop}%
\bibitem [{\citenamefont {Evans}\ and\ \citenamefont
  {Searles}(2002)}]{evans2002fluctuation}%
  \BibitemOpen
  \bibfield  {author} {\bibinfo {author} {\bibfnamefont {D.~J.}\ \bibnamefont
  {Evans}}\ and\ \bibinfo {author} {\bibfnamefont {D.~J.}\ \bibnamefont
  {Searles}},\ }\href@noop {} {\bibfield  {journal} {\bibinfo  {journal}
  {Advances in Physics}\ }\textbf {\bibinfo {volume} {51}},\ \bibinfo {pages}
  {1529} (\bibinfo {year} {2002})}\BibitemShut {NoStop}%
\bibitem [{\citenamefont {Gallavotti}\ and\ \citenamefont
  {Cohen}(1995)}]{gallavotti1995dynamical}%
  \BibitemOpen
  \bibfield  {author} {\bibinfo {author} {\bibfnamefont {G.}~\bibnamefont
  {Gallavotti}}\ and\ \bibinfo {author} {\bibfnamefont {E.~G.~D.}\ \bibnamefont
  {Cohen}},\ }\href@noop {} {\bibfield  {journal} {\bibinfo  {journal}
  {Physical review letters}\ }\textbf {\bibinfo {volume} {74}},\ \bibinfo
  {pages} {2694} (\bibinfo {year} {1995})}\BibitemShut {NoStop}%
\bibitem [{\citenamefont {Talkner}\ and\ \citenamefont
  {H{\"a}nggi}(2007)}]{talkner2007tasaki}%
  \BibitemOpen
  \bibfield  {author} {\bibinfo {author} {\bibfnamefont {P.}~\bibnamefont
  {Talkner}}\ and\ \bibinfo {author} {\bibfnamefont {P.}~\bibnamefont
  {H{\"a}nggi}},\ }\href@noop {} {\bibfield  {journal} {\bibinfo  {journal}
  {Journal of Physics A: Mathematical and Theoretical}\ }\textbf {\bibinfo
  {volume} {40}},\ \bibinfo {pages} {F569} (\bibinfo {year}
  {2007})}\BibitemShut {NoStop}%
\bibitem [{\citenamefont {Kurchan}(2000)}]{kurchan2000quantum}%
  \BibitemOpen
  \bibfield  {author} {\bibinfo {author} {\bibfnamefont {J.}~\bibnamefont
  {Kurchan}},\ }\href@noop {} {\bibfield  {journal} {\bibinfo  {journal} {arXiv
  preprint cond-mat/0007360}\ } (\bibinfo {year} {2000})}\BibitemShut {NoStop}%
\bibitem [{\citenamefont {Sagawa}\ and\ \citenamefont
  {Ueda}(2008)}]{sagawa2008second}%
  \BibitemOpen
  \bibfield  {author} {\bibinfo {author} {\bibfnamefont {T.}~\bibnamefont
  {Sagawa}}\ and\ \bibinfo {author} {\bibfnamefont {M.}~\bibnamefont {Ueda}},\
  }\href@noop {} {\bibfield  {journal} {\bibinfo  {journal} {Physical review
  letters}\ }\textbf {\bibinfo {volume} {100}},\ \bibinfo {pages} {080403}
  (\bibinfo {year} {2008})}\BibitemShut {NoStop}%
\bibitem [{\citenamefont {Sagawa}\ and\ \citenamefont
  {Ueda}(2010)}]{sagawa2010generalized}%
  \BibitemOpen
  \bibfield  {author} {\bibinfo {author} {\bibfnamefont {T.}~\bibnamefont
  {Sagawa}}\ and\ \bibinfo {author} {\bibfnamefont {M.}~\bibnamefont {Ueda}},\
  }\href@noop {} {\bibfield  {journal} {\bibinfo  {journal} {Physical review
  letters}\ }\textbf {\bibinfo {volume} {104}},\ \bibinfo {pages} {090602}
  (\bibinfo {year} {2010})}\BibitemShut {NoStop}%
\bibitem [{\citenamefont {Sagawa}\ and\ \citenamefont
  {Ueda}(2012)}]{sagawa2012nonequilibrium}%
  \BibitemOpen
  \bibfield  {author} {\bibinfo {author} {\bibfnamefont {T.}~\bibnamefont
  {Sagawa}}\ and\ \bibinfo {author} {\bibfnamefont {M.}~\bibnamefont {Ueda}},\
  }\href@noop {} {\bibfield  {journal} {\bibinfo  {journal} {Physical Review
  E}\ }\textbf {\bibinfo {volume} {85}},\ \bibinfo {pages} {021104} (\bibinfo
  {year} {2012})}\BibitemShut {NoStop}%
\bibitem [{\citenamefont {Callen}\ and\ \citenamefont
  {Welton}(1951)}]{callen1951irreversibility}%
  \BibitemOpen
  \bibfield  {author} {\bibinfo {author} {\bibfnamefont {H.~B.}\ \bibnamefont
  {Callen}}\ and\ \bibinfo {author} {\bibfnamefont {T.~A.}\ \bibnamefont
  {Welton}},\ }\href@noop {} {\bibfield  {journal} {\bibinfo  {journal}
  {Physical Review}\ }\textbf {\bibinfo {volume} {83}},\ \bibinfo {pages} {34}
  (\bibinfo {year} {1951})}\BibitemShut {NoStop}%
\bibitem [{\citenamefont {Seifert}(2012)}]{seifert2012stochastic}%
  \BibitemOpen
  \bibfield  {author} {\bibinfo {author} {\bibfnamefont {U.}~\bibnamefont
  {Seifert}},\ }\href@noop {} {\bibfield  {journal} {\bibinfo  {journal}
  {Reports on progress in physics}\ }\textbf {\bibinfo {volume} {75}},\
  \bibinfo {pages} {126001} (\bibinfo {year} {2012})}\BibitemShut {NoStop}%
\bibitem [{\citenamefont {Seifert}(2008)}]{seifert2008stochastic}%
  \BibitemOpen
  \bibfield  {author} {\bibinfo {author} {\bibfnamefont {U.}~\bibnamefont
  {Seifert}},\ }\href@noop {} {\bibfield  {journal} {\bibinfo  {journal} {The
  European Physical Journal B}\ }\textbf {\bibinfo {volume} {64}},\ \bibinfo
  {pages} {423} (\bibinfo {year} {2008})}\BibitemShut {NoStop}%
\bibitem [{\citenamefont {Ciliberto}(2017)}]{ciliberto2017experiments}%
  \BibitemOpen
  \bibfield  {author} {\bibinfo {author} {\bibfnamefont {S.}~\bibnamefont
  {Ciliberto}},\ }\href@noop {} {\bibfield  {journal} {\bibinfo  {journal}
  {Physical Review X}\ }\textbf {\bibinfo {volume} {7}},\ \bibinfo {pages}
  {021051} (\bibinfo {year} {2017})}\BibitemShut {NoStop}%
\bibitem [{\citenamefont {Bechhoefer}\ \emph {et~al.}(2020)\citenamefont
  {Bechhoefer}, \citenamefont {Ciliberto}, \citenamefont {Pigolotti},\ and\
  \citenamefont {Rold{\'a}n}}]{bechhoefer2020stochastic}%
  \BibitemOpen
  \bibfield  {author} {\bibinfo {author} {\bibfnamefont {J.}~\bibnamefont
  {Bechhoefer}}, \bibinfo {author} {\bibfnamefont {S.}~\bibnamefont
  {Ciliberto}}, \bibinfo {author} {\bibfnamefont {S.}~\bibnamefont
  {Pigolotti}}, \ and\ \bibinfo {author} {\bibfnamefont {E.}~\bibnamefont
  {Rold{\'a}n}},\ }\href@noop {} {\bibfield  {journal} {\bibinfo  {journal}
  {Journal of Statistical Mechanics: Theory and Experiment}\ }\textbf {\bibinfo
  {volume} {2020}},\ \bibinfo {pages} {064001} (\bibinfo {year}
  {2020})}\BibitemShut {NoStop}%
\bibitem [{\citenamefont {Seifert}(2019)}]{seifert2019stochastic}%
  \BibitemOpen
  \bibfield  {author} {\bibinfo {author} {\bibfnamefont {U.}~\bibnamefont
  {Seifert}},\ }\href@noop {} {\bibfield  {journal} {\bibinfo  {journal}
  {Annual Review of Condensed Matter Physics}\ }\textbf {\bibinfo {volume}
  {10}},\ \bibinfo {pages} {171} (\bibinfo {year} {2019})}\BibitemShut
  {NoStop}%
\bibitem [{\citenamefont {Jarzynski}(2011)}]{jarzynski2011equalities}%
  \BibitemOpen
  \bibfield  {author} {\bibinfo {author} {\bibfnamefont {C.}~\bibnamefont
  {Jarzynski}},\ }\href@noop {} {\bibfield  {journal} {\bibinfo  {journal}
  {Annu. Rev. Condens. Matter Phys.}\ }\textbf {\bibinfo {volume} {2}},\
  \bibinfo {pages} {329} (\bibinfo {year} {2011})}\BibitemShut {NoStop}%
\bibitem [{\citenamefont {Klages}\ \emph {et~al.}(2013)\citenamefont {Klages},
  \citenamefont {Just},\ and\ \citenamefont
  {Jarzynski}}]{klages2013nonequilibrium}%
  \BibitemOpen
  \bibfield  {author} {\bibinfo {author} {\bibfnamefont {R.}~\bibnamefont
  {Klages}}, \bibinfo {author} {\bibfnamefont {W.}~\bibnamefont {Just}}, \ and\
  \bibinfo {author} {\bibfnamefont {C.}~\bibnamefont {Jarzynski}},\ }\href@noop
  {} {\emph {\bibinfo {title} {Nonequilibrium statistical physics of small
  systems}}}\ (\bibinfo  {publisher} {Wiley Online Library},\ \bibinfo {year}
  {2013})\BibitemShut {NoStop}%
\bibitem [{\citenamefont {Shalizi}(2004)}]{shalizi2004backwards}%
  \BibitemOpen
  \bibfield  {author} {\bibinfo {author} {\bibfnamefont {C.~R.}\ \bibnamefont
  {Shalizi}},\ }\href@noop {} {\bibfield  {journal} {\bibinfo  {journal} {arXiv
  preprint cond-mat/0410063}\ } (\bibinfo {year} {2004})}\BibitemShut {NoStop}%
\bibitem [{\citenamefont {Jaynes}(1990)}]{jaynes1990probability}%
  \BibitemOpen
  \bibfield  {author} {\bibinfo {author} {\bibfnamefont {E.~T.}\ \bibnamefont
  {Jaynes}},\ }in\ \href@noop {} {\emph {\bibinfo {booktitle} {Maximum entropy
  and Bayesian methods}}}\ (\bibinfo  {publisher} {Springer},\ \bibinfo {year}
  {1990})\ pp.\ \bibinfo {pages} {1--16}\BibitemShut {NoStop}%
\bibitem [{\citenamefont {Kyburg}(1987)}]{kyburg1987basic}%
  \BibitemOpen
  \bibfield  {author} {\bibinfo {author} {\bibfnamefont {H.~E.}\ \bibnamefont
  {Kyburg}},\ }in\ \href@noop {} {\emph {\bibinfo {booktitle} {Advances in the
  Statistical Sciences: Foundations of Statistical Inference}}}\ (\bibinfo
  {publisher} {Springer},\ \bibinfo {year} {1987})\ pp.\ \bibinfo {pages}
  {219--232}\BibitemShut {NoStop}%
\bibitem [{\citenamefont {Sklar}(1995)}]{sklar1995physics}%
  \BibitemOpen
  \bibfield  {author} {\bibinfo {author} {\bibfnamefont {L.}~\bibnamefont
  {Sklar}},\ }\href@noop {} {\emph {\bibinfo {title} {Physics and chance:
  Philosophical issues in the foundations of statistical mechanics}}}\
  (\bibinfo  {publisher} {Cambridge University Press},\ \bibinfo {year}
  {1995})\BibitemShut {NoStop}%
\bibitem [{\citenamefont {Nowakowski}\ and\ \citenamefont
  {Lemarchand}(2002)}]{nowakowski2002thermal}%
  \BibitemOpen
  \bibfield  {author} {\bibinfo {author} {\bibfnamefont {B.}~\bibnamefont
  {Nowakowski}}\ and\ \bibinfo {author} {\bibfnamefont {A.}~\bibnamefont
  {Lemarchand}},\ }\href@noop {} {\bibfield  {journal} {\bibinfo  {journal}
  {Physica A: Statistical Mechanics and its Applications}\ }\textbf {\bibinfo
  {volume} {311}},\ \bibinfo {pages} {80} (\bibinfo {year} {2002})}\BibitemShut
  {NoStop}%
\bibitem [{\citenamefont {Kramers}(1940)}]{kramers1940brownian}%
  \BibitemOpen
  \bibfield  {author} {\bibinfo {author} {\bibfnamefont {H.~A.}\ \bibnamefont
  {Kramers}},\ }\href@noop {} {\bibfield  {journal} {\bibinfo  {journal}
  {Physica}\ }\textbf {\bibinfo {volume} {7}},\ \bibinfo {pages} {284}
  (\bibinfo {year} {1940})}\BibitemShut {NoStop}%
\bibitem [{\citenamefont {Van~Kampen}(1987)}]{van1987intrinsic}%
  \BibitemOpen
  \bibfield  {author} {\bibinfo {author} {\bibfnamefont {N.}~\bibnamefont
  {Van~Kampen}},\ }\href@noop {} {\bibfield  {journal} {\bibinfo  {journal}
  {Journal of statistical physics}\ }\textbf {\bibinfo {volume} {46}},\
  \bibinfo {pages} {933} (\bibinfo {year} {1987})}\BibitemShut {NoStop}%
\bibitem [{\citenamefont {Janssen}(1989)}]{janssen1989elimination}%
  \BibitemOpen
  \bibfield  {author} {\bibinfo {author} {\bibfnamefont {J.}~\bibnamefont
  {Janssen}},\ }\href@noop {} {\bibfield  {journal} {\bibinfo  {journal}
  {Journal of statistical physics}\ }\textbf {\bibinfo {volume} {57}},\
  \bibinfo {pages} {157} (\bibinfo {year} {1989})}\BibitemShut {NoStop}%
\bibitem [{\citenamefont {H{\"a}nggi}\ \emph {et~al.}(1990)\citenamefont
  {H{\"a}nggi}, \citenamefont {Talkner},\ and\ \citenamefont
  {Borkovec}}]{hanggi1990reaction}%
  \BibitemOpen
  \bibfield  {author} {\bibinfo {author} {\bibfnamefont {P.}~\bibnamefont
  {H{\"a}nggi}}, \bibinfo {author} {\bibfnamefont {P.}~\bibnamefont {Talkner}},
  \ and\ \bibinfo {author} {\bibfnamefont {M.}~\bibnamefont {Borkovec}},\
  }\href@noop {} {\bibfield  {journal} {\bibinfo  {journal} {Reviews of modern
  physics}\ }\textbf {\bibinfo {volume} {62}},\ \bibinfo {pages} {251}
  (\bibinfo {year} {1990})}\BibitemShut {NoStop}%
\bibitem [{\citenamefont {Ghonge}\ and\ \citenamefont
  {Vural}(2018)}]{ghonge2018temperature}%
  \BibitemOpen
  \bibfield  {author} {\bibinfo {author} {\bibfnamefont {S.}~\bibnamefont
  {Ghonge}}\ and\ \bibinfo {author} {\bibfnamefont {D.~C.}\ \bibnamefont
  {Vural}},\ }\href@noop {} {\bibfield  {journal} {\bibinfo  {journal} {Journal
  of Statistical Mechanics: Theory and Experiment}\ }\textbf {\bibinfo {volume}
  {2018}},\ \bibinfo {pages} {073102} (\bibinfo {year} {2018})}\BibitemShut
  {NoStop}%
\bibitem [{\citenamefont {Xu}(1996)}]{xu1996volume}%
  \BibitemOpen
  \bibfield  {author} {\bibinfo {author} {\bibfnamefont {L.}~\bibnamefont
  {Xu}},\ }\href@noop {} {\bibfield  {journal} {\bibinfo  {journal} {SIAM
  Review}\ }\textbf {\bibinfo {volume} {38}},\ \bibinfo {pages} {669} (\bibinfo
  {year} {1996})}\BibitemShut {NoStop}%
\bibitem [{\citenamefont {Swendsen}\ and\ \citenamefont
  {Wang}(2015)}]{swendsen2015gibbs}%
  \BibitemOpen
  \bibfield  {author} {\bibinfo {author} {\bibfnamefont {R.~H.}\ \bibnamefont
  {Swendsen}}\ and\ \bibinfo {author} {\bibfnamefont {J.-S.}\ \bibnamefont
  {Wang}},\ }\href@noop {} {\bibfield  {journal} {\bibinfo  {journal} {Physical
  Review E}\ }\textbf {\bibinfo {volume} {92}},\ \bibinfo {pages} {020103}
  (\bibinfo {year} {2015})}\BibitemShut {NoStop}%
\bibitem [{\citenamefont {Frenkel}\ and\ \citenamefont
  {Warren}(2015)}]{frenkel2015gibbs}%
  \BibitemOpen
  \bibfield  {author} {\bibinfo {author} {\bibfnamefont {D.}~\bibnamefont
  {Frenkel}}\ and\ \bibinfo {author} {\bibfnamefont {P.~B.}\ \bibnamefont
  {Warren}},\ }\href@noop {} {\bibfield  {journal} {\bibinfo  {journal}
  {American Journal of Physics}\ }\textbf {\bibinfo {volume} {83}},\ \bibinfo
  {pages} {163} (\bibinfo {year} {2015})}\BibitemShut {NoStop}%
\bibitem [{\citenamefont {Cerino}\ \emph {et~al.}(2015)\citenamefont {Cerino},
  \citenamefont {Puglisi},\ and\ \citenamefont
  {Vulpiani}}]{cerino2015consistent}%
  \BibitemOpen
  \bibfield  {author} {\bibinfo {author} {\bibfnamefont {L.}~\bibnamefont
  {Cerino}}, \bibinfo {author} {\bibfnamefont {A.}~\bibnamefont {Puglisi}}, \
  and\ \bibinfo {author} {\bibfnamefont {A.}~\bibnamefont {Vulpiani}},\
  }\href@noop {} {\bibfield  {journal} {\bibinfo  {journal} {Journal of
  Statistical Mechanics: Theory and Experiment}\ }\textbf {\bibinfo {volume}
  {2015}},\ \bibinfo {pages} {P12002} (\bibinfo {year} {2015})}\BibitemShut
  {NoStop}%
\bibitem [{\citenamefont {Procaccia}\ and\ \citenamefont
  {Levine}(1976)}]{procaccia1976potential}%
  \BibitemOpen
  \bibfield  {author} {\bibinfo {author} {\bibfnamefont {I.}~\bibnamefont
  {Procaccia}}\ and\ \bibinfo {author} {\bibfnamefont {R.}~\bibnamefont
  {Levine}},\ }\href@noop {} {\bibfield  {journal} {\bibinfo  {journal} {The
  Journal of Chemical Physics}\ }\textbf {\bibinfo {volume} {65}},\ \bibinfo
  {pages} {3357} (\bibinfo {year} {1976})}\BibitemShut {NoStop}%
\bibitem [{\citenamefont {Sagawa}\ and\ \citenamefont
  {Ueda}(2009)}]{sagawa2009minimal}%
  \BibitemOpen
  \bibfield  {author} {\bibinfo {author} {\bibfnamefont {T.}~\bibnamefont
  {Sagawa}}\ and\ \bibinfo {author} {\bibfnamefont {M.}~\bibnamefont {Ueda}},\
  }\href@noop {} {\bibfield  {journal} {\bibinfo  {journal} {Physical review
  letters}\ }\textbf {\bibinfo {volume} {102}},\ \bibinfo {pages} {250602}
  (\bibinfo {year} {2009})}\BibitemShut {NoStop}%
\bibitem [{\citenamefont {Pathria}(1983)}]{pathria1983phase}%
  \BibitemOpen
  \bibfield  {author} {\bibinfo {author} {\bibfnamefont {R.}~\bibnamefont
  {Pathria}},\ }\href@noop {} {\bibfield  {journal} {\bibinfo  {journal}
  {Canadian Journal of Physics}\ }\textbf {\bibinfo {volume} {61}},\ \bibinfo
  {pages} {228} (\bibinfo {year} {1983})}\BibitemShut {NoStop}%
\bibitem [{\citenamefont {Subrahmanyam}\ and\ \citenamefont
  {Barma}(1989)}]{subrahmanyam1989finite}%
  \BibitemOpen
  \bibfield  {author} {\bibinfo {author} {\bibfnamefont {V.}~\bibnamefont
  {Subrahmanyam}}\ and\ \bibinfo {author} {\bibfnamefont {M.}~\bibnamefont
  {Barma}},\ }\href@noop {} {\bibfield  {journal} {\bibinfo  {journal} {Journal
  of Physics A: Mathematical and General}\ }\textbf {\bibinfo {volume} {22}},\
  \bibinfo {pages} {L489} (\bibinfo {year} {1989})}\BibitemShut {NoStop}%
\bibitem [{\citenamefont {Dey}\ \emph {et~al.}(2020)\citenamefont {Dey},
  \citenamefont {Manchala}, \citenamefont {Basu}, \citenamefont {Banerjee},\
  and\ \citenamefont {Biswas}}]{dey2020finite}%
  \BibitemOpen
  \bibfield  {author} {\bibinfo {author} {\bibfnamefont {S.}~\bibnamefont
  {Dey}}, \bibinfo {author} {\bibfnamefont {P.}~\bibnamefont {Manchala}},
  \bibinfo {author} {\bibfnamefont {S.}~\bibnamefont {Basu}}, \bibinfo {author}
  {\bibfnamefont {D.}~\bibnamefont {Banerjee}}, \ and\ \bibinfo {author}
  {\bibfnamefont {S.}~\bibnamefont {Biswas}},\ }\href@noop {} {\bibfield
  {journal} {\bibinfo  {journal} {Physica Scripta}\ } (\bibinfo {year}
  {2020})}\BibitemShut {NoStop}%
\bibitem [{\citenamefont {Bhattacharyya}\ and\ \citenamefont
  {Bhattacharjee}(1998)}]{bhattacharyya1998finite}%
  \BibitemOpen
  \bibfield  {author} {\bibinfo {author} {\bibfnamefont {S.}~\bibnamefont
  {Bhattacharyya}}\ and\ \bibinfo {author} {\bibfnamefont {J.}~\bibnamefont
  {Bhattacharjee}},\ }\href@noop {} {\bibfield  {journal} {\bibinfo  {journal}
  {EPL (Europhysics Letters)}\ }\textbf {\bibinfo {volume} {43}},\ \bibinfo
  {pages} {129} (\bibinfo {year} {1998})}\BibitemShut {NoStop}%
\bibitem [{\citenamefont {Danchev}\ \emph {et~al.}(2000)\citenamefont
  {Danchev}, \citenamefont {Tonchev} \emph {et~al.}}]{danchev2000theory}%
  \BibitemOpen
  \bibfield  {author} {\bibinfo {author} {\bibfnamefont {D.~M.}\ \bibnamefont
  {Danchev}}, \bibinfo {author} {\bibfnamefont {N.~S.}\ \bibnamefont
  {Tonchev}},  \emph {et~al.},\ }\href@noop {} {\emph {\bibinfo {title} {Theory
  of critical phenomena in finite-size systems: scaling and quantum
  effects}}},\ Vol.~\bibinfo {volume} {9}\ (\bibinfo  {publisher} {World
  Scientific},\ \bibinfo {year} {2000})\BibitemShut {NoStop}%
\bibitem [{\citenamefont {Jaynes}(1957{\natexlab{a}})}]{jaynes1957information}%
  \BibitemOpen
  \bibfield  {author} {\bibinfo {author} {\bibfnamefont {E.~T.}\ \bibnamefont
  {Jaynes}},\ }\href@noop {} {\bibfield  {journal} {\bibinfo  {journal}
  {Physical review}\ }\textbf {\bibinfo {volume} {106}},\ \bibinfo {pages}
  {620} (\bibinfo {year} {1957}{\natexlab{a}})}\BibitemShut {NoStop}%
\bibitem [{\citenamefont
  {Jaynes}(1957{\natexlab{b}})}]{jaynes1957information2}%
  \BibitemOpen
  \bibfield  {author} {\bibinfo {author} {\bibfnamefont {E.~T.}\ \bibnamefont
  {Jaynes}},\ }\href@noop {} {\bibfield  {journal} {\bibinfo  {journal}
  {Physical review}\ }\textbf {\bibinfo {volume} {108}},\ \bibinfo {pages}
  {171} (\bibinfo {year} {1957}{\natexlab{b}})}\BibitemShut {NoStop}%
\bibitem [{\citenamefont {Jaynes}(1968)}]{jaynes1968prior}%
  \BibitemOpen
  \bibfield  {author} {\bibinfo {author} {\bibfnamefont {E.~T.}\ \bibnamefont
  {Jaynes}},\ }\href@noop {} {\bibfield  {journal} {\bibinfo  {journal} {IEEE
  Transactions on systems science and cybernetics}\ }\textbf {\bibinfo {volume}
  {4}},\ \bibinfo {pages} {227} (\bibinfo {year} {1968})}\BibitemShut {NoStop}%
\bibitem [{\citenamefont {Jaynes}(1988)}]{jaynes1988relation}%
  \BibitemOpen
  \bibfield  {author} {\bibinfo {author} {\bibfnamefont {E.~T.}\ \bibnamefont
  {Jaynes}},\ }in\ \href@noop {} {\emph {\bibinfo {booktitle} {Maximum-entropy
  and Bayesian methods in science and engineering}}}\ (\bibinfo  {publisher}
  {Springer},\ \bibinfo {year} {1988})\ pp.\ \bibinfo {pages}
  {25--29}\BibitemShut {NoStop}%
\bibitem [{\citenamefont {Casas-V{\'a}zquez}\ and\ \citenamefont
  {Jou}(2003)}]{casas2003temperature}%
  \BibitemOpen
  \bibfield  {author} {\bibinfo {author} {\bibfnamefont {J.}~\bibnamefont
  {Casas-V{\'a}zquez}}\ and\ \bibinfo {author} {\bibfnamefont {D.}~\bibnamefont
  {Jou}},\ }\href@noop {} {\bibfield  {journal} {\bibinfo  {journal} {Reports
  on Progress in Physics}\ }\textbf {\bibinfo {volume} {66}},\ \bibinfo {pages}
  {1937} (\bibinfo {year} {2003})}\BibitemShut {NoStop}%
\bibitem [{\citenamefont {Hartmann}\ \emph
  {et~al.}(2004{\natexlab{b}})\citenamefont {Hartmann}, \citenamefont
  {Mahler},\ and\ \citenamefont {Hess}}]{hartmann2004existence}%
  \BibitemOpen
  \bibfield  {author} {\bibinfo {author} {\bibfnamefont {M.}~\bibnamefont
  {Hartmann}}, \bibinfo {author} {\bibfnamefont {G.}~\bibnamefont {Mahler}}, \
  and\ \bibinfo {author} {\bibfnamefont {O.}~\bibnamefont {Hess}},\ }\href@noop
  {} {\bibfield  {journal} {\bibinfo  {journal} {Physical review letters}\
  }\textbf {\bibinfo {volume} {93}},\ \bibinfo {pages} {080402} (\bibinfo
  {year} {2004}{\natexlab{b}})}\BibitemShut {NoStop}%
\bibitem [{\citenamefont {Hartmann}\ and\ \citenamefont
  {Mahler}(2005)}]{hartmann2005measurable}%
  \BibitemOpen
  \bibfield  {author} {\bibinfo {author} {\bibfnamefont {M.}~\bibnamefont
  {Hartmann}}\ and\ \bibinfo {author} {\bibfnamefont {G.}~\bibnamefont
  {Mahler}},\ }\href@noop {} {\bibfield  {journal} {\bibinfo  {journal} {EPL
  (Europhysics Letters)}\ }\textbf {\bibinfo {volume} {70}},\ \bibinfo {pages}
  {579} (\bibinfo {year} {2005})}\BibitemShut {NoStop}%
\end{thebibliography}


\end{document}